\documentclass[lettersize,journal]{IEEEtran}

\usepackage{amsmath,amsfonts}
\usepackage{algorithmic}
\usepackage{algorithm}
\usepackage{amsmath}
\usepackage{array}
\usepackage[caption=false,font=normalsize,labelfont=sf,textfont=sc]{subfig}
\usepackage{caption}
\usepackage{subcaption}
\usepackage{textcomp}
\usepackage{stfloats}
\usepackage{multirow} 
\usepackage{booktabs}
\usepackage{url}
\usepackage{verbatim}
\usepackage{siunitx}
\usepackage{graphicx}
\usepackage{cite}
\usepackage{siunitx}
\usepackage{xcolor}
\IEEEoverridecommandlockouts
\begin{document}

\title{SimLOB: Learning Representations of Limit Order Book for Financial Market Simulation}

\author{Yuanzhe~Li\IEEEauthorrefmark{1},
        Yue~Wu\IEEEauthorrefmark{1},
        Muyao~Zhong,
        Shengcai~Liu,
        and Peng~Yang\IEEEauthorrefmark{2}%
        
\thanks{Yuanzhe Li is with the Department of Statistics and Data Science, Southern University of Science and Technology, Shenzhen 518055, China (e-mail: 12113016@mail.sustech.edu.cn).}%

\thanks{Yue Wu is with the Department of Statistics and Data Science, Southern University of Science and Technology, Shenzhen 518055, China (e-mail: wuy2021@mail.sustech.edu.cn).}%

\thanks{Muyao Zhong is with the Department of Computer Science and Engineering, Southern University of Science and Technology, Shenzhen 518055, China (e-mail: 12149032@mail.sustech.edu.cn).}%

\thanks{ Shengcai Liu is with the Department of Computer Science and Engineering, Southern University of Science and Technology, Shenzhen 518055, China (e-mail: liusc3@sustech.edu.cn).}%

\thanks{Peng Yang is with the Department of Statistics and Data Science, the Department of Computer Science and Engineering, and the Guangdong Provincial Key Laboratory of Brain-Inspired Intelligent Computation, Southern University of Science and Technology, Shenzhen 518055, China (e-mail: yangp@sustech.edu.cn).}%

\thanks{\IEEEauthorrefmark{1} Contributed equally to this work. Authors are listed alphabetically. }

\thanks{\IEEEauthorrefmark{2} Corresponding author.}
}

\maketitle

\begin{abstract}
Financial market simulation (FMS) serves as a promising tool for understanding market anomalies and underlying trading behaviors. To ensure high-fidelity simulations, it is crucial to calibrate the FMS model to generate data closely resembling the observed market data. Previous efforts primarily focused on calibrating merely the mid-price data, leading to essential information loss of the market activities and thus biasing the calibrated model. The Limit Order Book data (LOB) is the fundamental data that fully captures the market microstructure and is adopted by exchanges worldwide. However, LOB is not applicable to existing calibration objective functions due to its unique properties, including high dimensionality, differences in feature scales, and cross-feature constraints. This paper proposes an efficient Transformer-based autoencoder to explicitly learn the vectorized representations of LOB, addressing its unique challenges. The resulted latent vector, which captures the major information of LOB, can then be applied for calibration. Extensive experiments show that the learned latent representation preserves not only the non-linear auto-correlation in the temporal axis but also the precedence between successive price levels of LOB. Besides, we verify that the representation learning aligns with the downstream calibration tasks, making the first advancement in calibrating FMS on LOB. 
\end{abstract}



\section{Introduction}
One important and challenging issue in mining financial market data is to discover the causes of the abnormal market phenomena, e.g., flash crash \cite{menkveld2019flash,paulin2019understanding} and spoofing \cite{Wang2017spoofing,spoofing_2}.
Traditional machine learning methods are well-equipped to detect the anomalies but lack of the financial interpretability of the causes \cite{yuAgentbasedFrameworkPolicy2023,zhao2025qfr}. 
Given observed market time series data, financial market simulation (FMS) aims to directly approximate its underlying data generating process, informed by the trading rules of the real market \cite{kellSystematicLiteratureReview2022}. 
By simulating the underlying trading activities behind the observed market data, FMS is able to discover how the structure of the financial market changed over time and explain the changes on the micro level \cite{fatrasAgentbasedSimulationEvaluating2023,hofman2021integrating}.  

Generally, FMS models the market as a parameterized multi-agent system, $M(\mathbf{w})$, simulating various types of traders and the exchange as distinct agents, where $M$ represents the FMS simulator and $\mathbf{w}$ serves as a vector of its control parameters. 
Each trading agent (TA) interacts with the exchange agent (EA) who implements the real matchmaking rules of the market. 
Since the 1990s, diverse trading agents have been designed to simulate chartists, fundamentalists, momentum traders, high-frequency traders, and so on \cite{axtell2022agent}, capturing most of the trading behaviors observed in real markets. 
When FMS is running, TAs continuously submit orders to the EA, which then produces the simulated market data by matching these orders. 
Normally, the exchange will publish the newest market data $\hat{\mathbf{x}}(t)$ at a certain frequency (e.g., 1 second). 
Hence, the FMS can be viewed as a market data generative process, denoted as $M(\mathbf{w})=\mathbf{X}^\mathbf{w}_T=\{\mathbf{x}^\mathbf{w}(t)\}_{t=1}^T$, where $T \in \mathbb{N}^+$ denotes the data sequence length. 

While FMSs have demonstrated the capability to simulate general properties of financial market data, market participants, in practice,  often focus on specific time periods for purposes such as anomaly detection, investment advice, or counterfactual reasoning\cite{calibrate_for}. Thus, FMS must accurately capture the underlying market processes during these targeted periods..
To achieve this, FMS aims to simulate the financial market data, $\hat{\mathbf{X}}_T = \{\hat{\mathbf{x}}(t)\}_{t=1}^T$, such that the simulated data $\mathbf{X}^\mathbf{w}_T$ closely resembles the observed data $\hat{\mathbf{X}}_T$. This alignment compels  $M(\mathbf{w})$ to approximate the data generative process of $\hat{\mathbf{X}}_T$. 
This process largely relies on tuning the parameters $\mathbf{w}$ of $M(\mathbf{w})$ to minimize the discrepancy between the two data, denoted as $D(\hat{\mathbf{X}}_T, M(\mathbf{w}))=D(\hat{\mathbf{X}}_T, \mathbf{X}^\mathbf{w}_T)$, a procedure also known as calibration\cite{bridging2022ye}. 
Unfortunately, this task is challenging, as the simulated data exhibits highly non-linearity due to the complex, rule-based interactions among diverse agents \cite{fabretti2013problem}. 

Advanced methods from both the fields of optimization and statistical inference have been developed to address the calibration problem \cite{wang2024alleviating,platt2020comparison,yang2025towards}.
Optimization methods primarily focus on  minimizing the discrepancy through black-box optimization techniques \cite{bai2022efficient}, while statistical inference approaches estimate the likelihood or posterior distribution of $\hat{\mathbf{X}}_T$ and $\mathbf{w}$ for selected samples subject to $D(\hat{\mathbf{X}}_T, M(\mathbf{w})) < \epsilon$ \cite{grazzini2017bayesian,bayesian2023ieee}.
In both ways, the calibration is conducted in an iterative randomized sampling manner in the parameter space and the sampling is mostly guided by the discrepancy. 
Intuitively, the more information observed from the market included in $\hat{\mathbf{X}}_T$, the closer $M(\mathbf{w})$ is expected to approximate the underlying data generative process. 
Most FMS works merely consider the mid-price data as $\hat{\mathbf{X}}_T$ \cite{platt2020comparison,yang2024evolutionary}, other important information of observable data, e.g., trading volume, bid/ask directions, and inter-arrival time of orders. 
Consequently, the calibrated $M(\mathbf{w})$ fails to capture the underlying market structure.

Compared to mid-price data, the Limit Order Book data (LOB) provides a more detailed and informative source of market information. It offers a comprehensive view of all untraded orders, capturing critical market dynamics beyond simple price changes \cite{LOB_book}. As the fundamental market data, LOB is widely adopted by most world-class securities exchanges \cite{kozhan2012information,yang2024reducing}. 
In LOB, all the untraded orders fall into either the ask side (for selling) or the bid side (for buying) according to their order directions. 
When a new order arrives, the LOB is updated immediately in one of the two ways: if the order price matches a price on the opposite side, the order is executed and the corresponding orders on the opposite side are removed; otherwise, the new order is added to its respective side. On each side, the untraded orders are organized in the "price-first-time-second" manner \cite{chiarella2002simulation}. 
Notably, the untraded orders in LOB implicitly reflect the trading intentions of the entire market. 
Besides, the immediate updates to the LOB provide a highly fine-grained view of market microstructure. 
On this basis, simulating LOB offers a conceptually close approximation to the real data generative process underlying the market scenario of interest.

\begin{figure}[!t]
\centering
\includegraphics[width=3.5in]{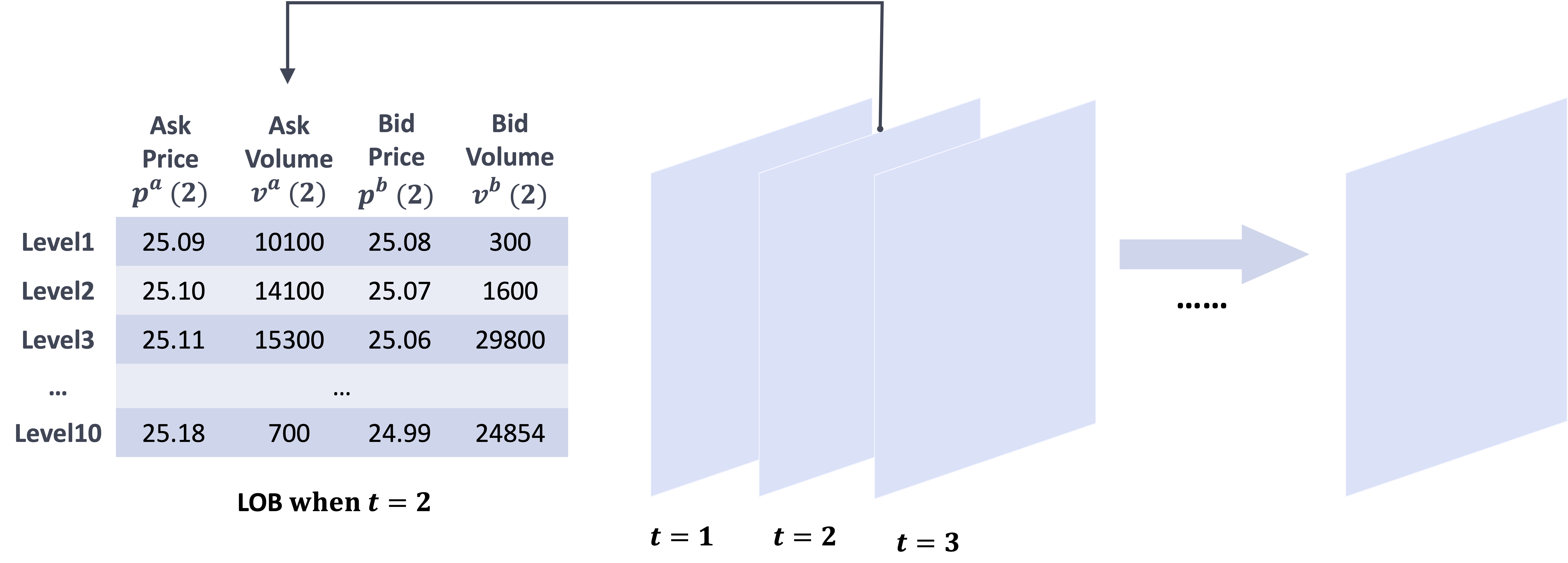}
\caption{An illustration of the LOB structure.}
\label{fig:lob data}
\end{figure}

Unfortunately, it is not straightforward to apply LOB into existing discrepancy measures. Representative discrepancies like Euclidean distances \cite{recchioni2015calibration,grazzini2015estimation} and probabilistic discrepancy\cite{bai2022efficient} all require $\hat{\mathbf{X}}_T$ and $\mathbf{X}^\mathbf{w}_T$ to be vectorized inputs, while the $T$ time steps LOB is publicized in the form of normally a $10 \times 4 \times T$ matrix data. That is, at each time step $t$, the best 10 price levels on both sides of LOB as well as the associated total volume of the untraded orders are output as $\hat{\mathbf{x}}(t)  \in \mathbb{R}^{10 \times 4}$. More specifically, $\hat{\mathbf{x}}(t)=[[p^b_1(t),v^b_1(t),p^a_1(t),v^a_1(t)];…;[p^b_{10}(t),v^b_{10}(t),p^a_{10}(t),v^a_{10}(t)]]$, where $p_i$ and $v_i$ denote the price level and total volume for the $i$-th position, and $a$ and $b$ indicate the ask and bid sides, respectively (see Fig. \ref{fig:lob data})., i.e., $\hat{mp}(t)=\frac{p^a_1(t)+p^b_1(t)}{2} \in \mathbb{R}^1$, which naturally lose much important information of the underlying data generative process. 

To our best knowledge, this is the first work attempting to calibrate FMS with respect to LOB. The key idea is to propose an autoencoder framework to learn a low-dimensional, vectorized representation (the latent vector) of LOB, enabling the application of well-established discrepancy measures for calibration. 
Specifically, if the output of the decoder closely resembles the input of the encoder, the latent vector is believed to be an effective representation of the original LOB. 
The main challenges stem from the unique properties of LOB, which involves not only the non-linear auto-correlation in the temporal axis but also the precedence between successive price levels at each time step, which may not be commonly seen in other time series data. 
Although recent studies have increasingly used neural networks to process LOB, they have not applied the autoencoder framework or explicitly studied the vectorized representations. Hence, this work offers the first solution to this problem and is expected to be further appreciated by both the neural network and quantitative finance fields. 

To capture the aforementioned unique properties in the latent vector, the proposed encoder and the symmetric decoder each contain three blocks: a fully connected network for extracting price precedence features, a Transformer stack for capturing the temporal auto-correlation, and a final fully connected network for dimension reduction. 
Four groups of experiments on synthetic and real data are conducted.
Key findings include: 1) The proposed method significantly outperforms existing LOB-oriented neural networks; 2) Calibrating LOB leads to better FMS than traditional mid-price calibration; 3) The model effectively learns vectorized LOB representations, whereas  convolutional layers in previous works are less effective; 4) The better representation of LOB, the better calibration of FMS.  

The rest of this paper is organized as follows. Section II reviews related works on FMS calibration and deep learning for LOB. Section III describes the problem definition, challenges, and the proposed LOB autoencoder. Section IV reports the empirical findings, and Section V concludes the paper.

\section{Related Work}
\subsection{Calibration of FMS} 
Traditional FMS works focus on designing rule-based agents to imitate various trading behaviors. 
While FMSs have reproduced and explained various common phenomena across stocks\cite{abm-gen2020ieee,vyetrenko2020get}, they received increasing criticisms for simulating only macro properties of the market than specifically targeted periods \cite{platt2018can,abm-blame2020ieee}.    
To address these criticisms, recent efforts have focused on developing effective calibration objectives to improve the fidelity of FMS \cite{platt2020comparison}. Since 2015, several calibration objectives have been proposed.  
A straightforward way is to simply calculate the MSE between the simulated mid-price vector and the observed mid-price vector \cite{recchioni2015calibration}. 
Methods of Simulated Moments calculate the statistical moments of both the observed and simulated mid-price and measure the distances between the two vectors of moments as the discrepancy \cite{grazzini2015estimation}. 
To better utilize the temporal properties of the mid-price vector, some information-theoretic criteria-based discrepancies are proposed based on various time windows \cite{barde2017practical}.  
Francesco \cite{lamperti2018information} tries to compare the distribution distance between the simulated and observed mid-price data.
The Kolmogorov-Smirnov (K-S) test is generalized to multi-variate data, while their empirical verification was only conducted on the mid-price vector and the traded volume vector \cite{bai2022efficient}.
To summarize, existing calibration functions all require the vectorized input format and related works mainly calibrate to the mid-price data.  

\subsection{Deep Learning for LOB} 
The field of deep learning for LOB have not been fully exploited.
Most of the studied tasks are to predict the price movement \cite{prata2024lob,jain2024limitorderbooksimulations}, a classification problem.
Since classification with neural networks is known to be effective, the end-to-end solutions are naturally considered for predicting the price movement from LOB, and the vectorized representation learning is not explicitly defined.
On the contrary, since the FMS models are mostly rule-based (especially the matchmaking rule of updating the LOB), they are non-differentiable and cannot be trained end-to-end with the feature extraction layers.
Thus, the vectorized representation learning needs special treatments in FMS.

Recent works propose to generate LOB data with deep learning, which is quite close to the goal of FMS \cite{assefa2020generating}.
However, they do not involve any validation or calibration steps to force the generated data resembling any given observed LOB but are only limited to following some macro properties of the market \cite{coletta2021towards,coletta2022learning,Coletta2023}. Besides, those deep generative models lack the interpretability of the quotes and trades in the real market.
Some of the models also are not informed by the real-world matchmaking rules for updating LOB \cite{li2020generating,cao2022synthetic}, and all of them are not explainable in why the specific orders are submitted.
These issues impose significant restrictions on the use of these models as the simulators since one can hardly intervene in the order streams to do a "what-if" test and reveal the micro-structured causes of certain financial events \cite{darley2007nasdaq,hamill2015agent}.
Furthermore, they explicitly require the ground truth order streams as input to train the networks, which is not available in the setting of FMS calibration problems.

Among them, the convolution layers and Long-Short Term Memory (LSTM) are the most used architecture.
Recently, the Transformer model has been adopted instead of LSTM, but the convolution layers are still kept \cite{yang2022ocet,wallbridge2020transformers}.
This work argues for abandoning the convolution layer as it is empirically found less effective in learning from LOB.

\section{The Proposals}
\subsection{The Problem Definition and Challenges}

To feed the LOB to the well-established calibration functions $D$, which only accept the vectorized input, we propose to learn a function $f$ that can represent the time series LOB data in a vectorized latent space. 
The requirement of $f$ is two-fold: first, $f$ naturally constitutes a dimension reduction process, mapping a $\tau$-step LOB $\hat{\mathbf{X}}_\tau \in \mathbb{R}^{10\times 4\times \tau}$ to $f(\hat{\mathbf{X}}_\tau) \in \mathbb{R}^{1\times \tilde{\tau}}$ with $\tilde{\tau} \in \mathbb{N}^+$ and $\tilde{\tau} \ll 40\tau$; second, within the latent space of $\mathbb{R}^{1\times \tilde{\tau}}$, the information of LOB can be largely preserved for downstream calibration tasks. 
One verification is the existence of an inverse function $g: \mathbb{R}^{1\times \tilde{\tau}} \rightarrow \mathbb{R}^{10\times 4\times \tau}$ such that $g(f(\hat{\mathbf{X}}_\tau))$ resembles the original $\hat{\mathbf{X}}_\tau$. 
To this end, the autoencoder architecture is considered, where $f$ constitutes the encoder network and $g$ is modeled by the decoder. 
The latent vector $\mathbf{Z}\in \mathbb{R}^{1\times \tilde{\tau}}$ between the encoder and the decoder is the learned representation of $\hat{\mathbf{X}}_\tau$, i.e., $\mathbf{Z}=f(\hat{\mathbf{X}}_\tau)$, and $g(\mathbf{Z})$ is the reconstructed data. Generally, the problem of learning representations for LOB can be formally defined as follows:

\begin{align}
\min_{g,f} D(\hat{\mathbf{X}}_\tau, g(f(\hat{\mathbf{X}}_\tau))).
\label{eq: learning representations for LOB}
\end{align}

The challenges of this representation learning problem lie in the specific properties of LOB, which can be generally stated as follows. 
\begin{enumerate}
    \item \textbf{Data at successive time steps is highly non-linearly auto-correlated.} The LOB data at $t + 1$ time step is constructed by matchmaking the LOB at the $t$ time step with all incoming orders between $t$ and $t + 1$. This matchmaking process typically involves complex if-else rules. Besides, theses incoming orders are generated by traders with various strategies based on the historical LOB, especially the LOB at $t$ time step. More formally, let $\hat{\mathbf{x}}(t)$ be the LOB at $t$ time step, the LOB at $t + 1$ can then be constructed as 
    $\hat{\mathbf{x}}(t+1) = \textbf{Matchmake}\left(\hat{\mathbf{x}}(t), \left\{\text{Trader}_i(\hat{\mathbf{x}}(t))\right\}_{i=1}^N \right)$
    where $N$ traders, each denoted as $\text{Trader}_i$, submits one order based on $\hat{\mathbf{x}}(t)$. Clearly, this construction results in a non-linearly auto-correlated time series.

    \item \textbf{Data has constraints on complex precedence between successive price columns at each time step.} For price columns, the following constraints apply: (1) ask prices must exceed bid prices, i.e., 
    $p^a_i(t) > p^b_j(t),  1 \leq i,j \leq 10$; (2) 
    Ask prices at lower levels (further from the best) should be smaller than those at higher levels, i.e.,
    $p^a_i(t) > p^a_j(t),  1 \leq i<j \leq 10$; (3) 
    Bid prices at lower levels should be larger than those at higher levels, i.e., 
    $p^b_i(t) < p^b_j(t),  1 \leq i<j \leq 10$.  No such constraints apply to volume columns.
    
    \item \textbf{Data is with different scales.} Prices typically fluctuate around several thousands of cents with a tick size of 1 cent, while volumes range from 100 or 1000 shares to millions of shares, dependent on the rules of exchanges.

\end{enumerate}

To summarize, learning representations of LOB requires jointly tackling the aforementioned three challenges. While the challenges of multi-scale and non-linearly auto-correlation have been extensively investigated in the literature, the precedence constraints have been rarely seen in other fields. Further more, it is uncommon to address them simultaneously with one representation model. These difficulties jointly offer LOB unique challenges and a good arena for studying representation learning of such type of data, particularly as the reconstructed multi-dimensionality data must maintain precedence constraints and avoid "intersection" at each time step.

\begin{figure*}[!t]
\centering
\includegraphics[width=0.9\textwidth, height=0.36\textwidth]{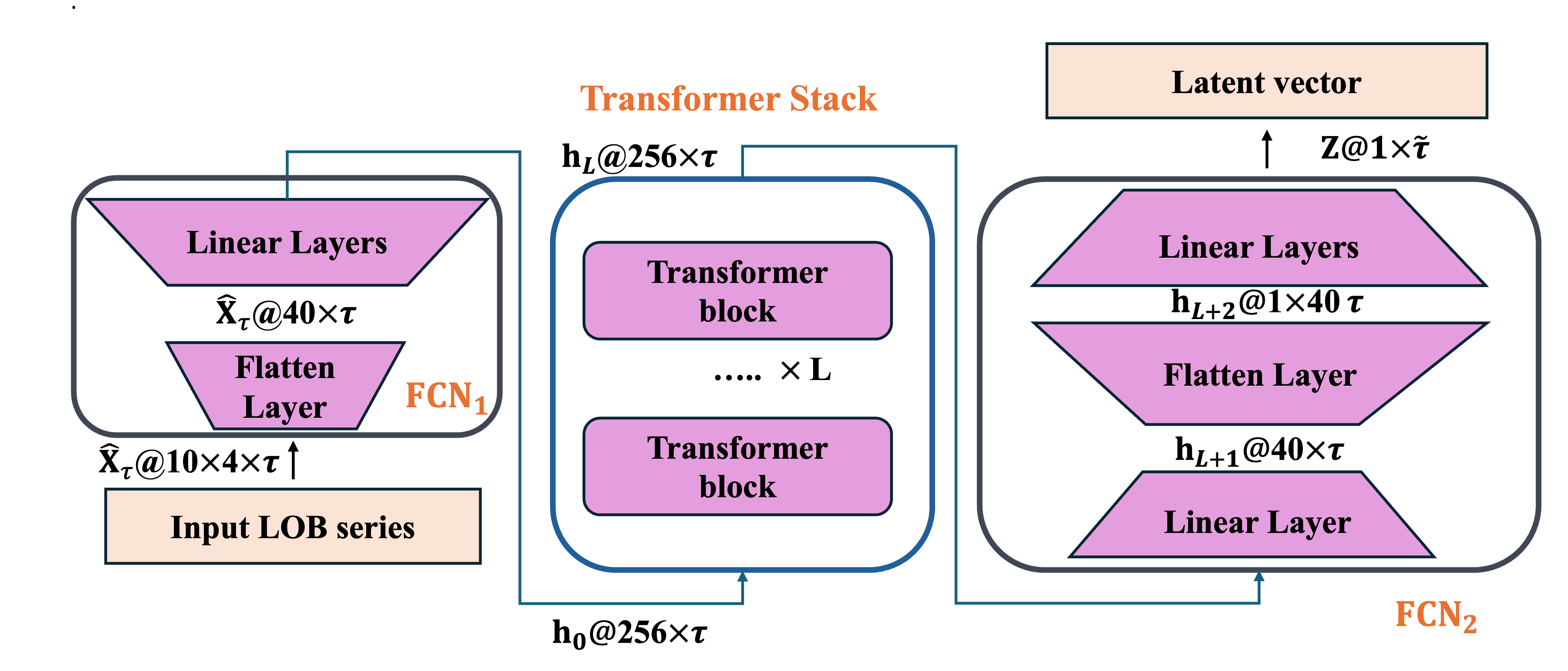}
\caption{Architecture of the proposed encoder, comprising a fully connected layer for feature extraction, a transformer stack, and another fully connected layer for dimension reduction.}
\label{fig: structure of model}
\end{figure*}

\subsection{The Network Architecture}


As described above, the encoder network should be able to effectively handle both the non-linear auto-correlation and the precedence constraints of price levels.
On this basis, the proposed encoder consists of three components: a feature extraction block ($\mathrm{FCN}_1$), a Transformer stack, and a dimension reduction block ($\mathrm{FCN}_2$). 
The architecture of the encoder is depicted in Fig. \ref{fig: structure of model}, with details provided below.

\textbf{The Feature Extraction Block.} The $\mathrm{FCN}_1$ is designed for extracting useful features from LOB. 
Previous machine learning methods for LOB often rely on pre-processed hand-crafted features to describe market dynamics \cite{ntakaris2018benchmark}. 
Those works typically consider the price levels as independent features and seldom deal with the precedence relationship. 
Recent pioneer work \cite{zhang2019deeplob} uses convolution neural networks (CNNs) to automatically extract the features from LOB.
The intuition is that CNN may capture the precedence between the price levels through convolutions like what has been done to the pixels.
However, the 10 levels of bid/ask prices and volumes have quite different scales and meanings, making effective convolution challenging.

This work utilizes a fully connected network (FCN) for feature extraction by $\mathbf{h}_0=\mathrm{FCN}_1(\hat{\mathbf{X}}_\tau)$.
First, the LOB at each time step $\hat{\mathbf{x}}(t) \in \mathbb{R}^{10 \times 4}$ is flattened into a vector $\mathbb{R}^{40 \times 1}$ using a linear layer.
Then the times series $\hat{\mathbf{X}}_\tau \in \mathbb{R}^{40 \times \tau}$ is projected into higher-dimensional space $\mathbf{h}_0 \in \mathbb{R}^{256 \times \tau}$ to capture richer feature representations. The intuition behind this is that traditional financial studies often construct hundreds of hand-crafted features from LOB that work well, and the $\mathrm{FCN}_1$ mimics this manual process via deep learning to automatically generate hundreds of non-linear features.

\textbf{The Transformer Block.} The second component of the encoder is a Transformer stack with $L$ vanilla Transformers \cite{vaswani2017attention}.
It aims to learn the non-linear auto-correlation between successive time steps using the multi-headed self-attention (MSA), feed-forward layers, and layer normalization (LN).
Specifically, at each $l$-th Transformer, the computations are: 
\begin{align*}
\mathbf{h}_l^{\prime}&=\mathrm{MSA}(\mathrm{LN}(\mathbf{h}_{l-1}))+\mathbf{h}_{l-1}, \quad &l = 1, \ldots, L \\
\mathbf{h}_l&=\mathrm{FCN}(\mathrm{LN}(\mathbf{h}_l^{\prime}))+\mathbf{h}_l^{\prime}, \quad &l = 1, \ldots, L
\end{align*}
where $\mathbf{h}_L$ is the output of the Transformer stack and the input of $\mathrm{FCN}_2$. 


\textbf{The Dimension Reduction Block.} Note that the Transformer does not change the shape of data, $\mathbf{h}_l \in \mathbb{R}^{256 \times \tau}$. Therefore,
$\mathrm{FCN}_2$ is designed to reduce the dimensionality to obtain the latent vector $\mathbf{Z} \in \mathbb{R}^{1\times \tilde{\tau}}$.
For that purpose, $\mathbf{h}_L$ is first projected to $\mathbf{h}_{L+1} \in \mathbb{R}^{40 \times \tau}$ with a linear layer. 
$\mathbf{h}_{L+1}$ is then flattened into $\mathbf{h}_{L+2} \in \mathbb{R}^{1 \times 40\tau}$ by concatenating its rows.
At last, three linear layers further reduce $\mathbf{h}_{L+2}$ to $\mathbf{Z}=\{z(t)\}^{\tilde{\tau}}_{t=1}$. 

\textbf{The Decoder.} The decoder is symmetric to the encoder. 
The latent vector $\mathbf{Z}$ is first expanded to  $\mathbb{R}^{256 \times \tau}$ via an inverted $\mathrm{FCN_2}$, followed by a stack of $L$ vanilla Transformers. Finally, an inverted  $\mathrm{FCN_1}$ reconstructs the data as $\mathbf{X}^r_\tau=\{\mathbf{x}^r(t)\}^\tau_{t=1}=g(f(\hat{\mathbf{X}}_\tau)) \in \mathbb{R}^{10\times 4\times \tau}$. 

\textbf{Implementation Details.} The proposed autoencoder, \textbf{SimLOB}, processes observed LOB by spliting it into multiple segments of $\tau=100$ time steps, as suggested by \cite{prata2024lob}.
Through sensitive analysis presented in Section \ref{sec:group4_sensitivity}, we set $L=2$ and $\tilde{\tau}=128$ by default.
The reconstruction error is defined as: 
\begin{align*}
    \mathbf{Err}_r=\frac{1}{4000}\sum^{10}_i\sum^4_j\sum^{100}_t (\hat{\mathbf{x}}_{i,j}(t) - \mathbf{x}^r_{i,j}(t))^2
\end{align*}
and the training loss $\mathbf{Loss}_r$ averages the reconstruction errors across $M=128$ training samples.
SimLOB is trained by Adam \cite{kingma2014adam} with a learning rate of 1E-4 for 200 epochs.

\begin{figure*}[t]
\centering
\includegraphics[width=0.65\textwidth]{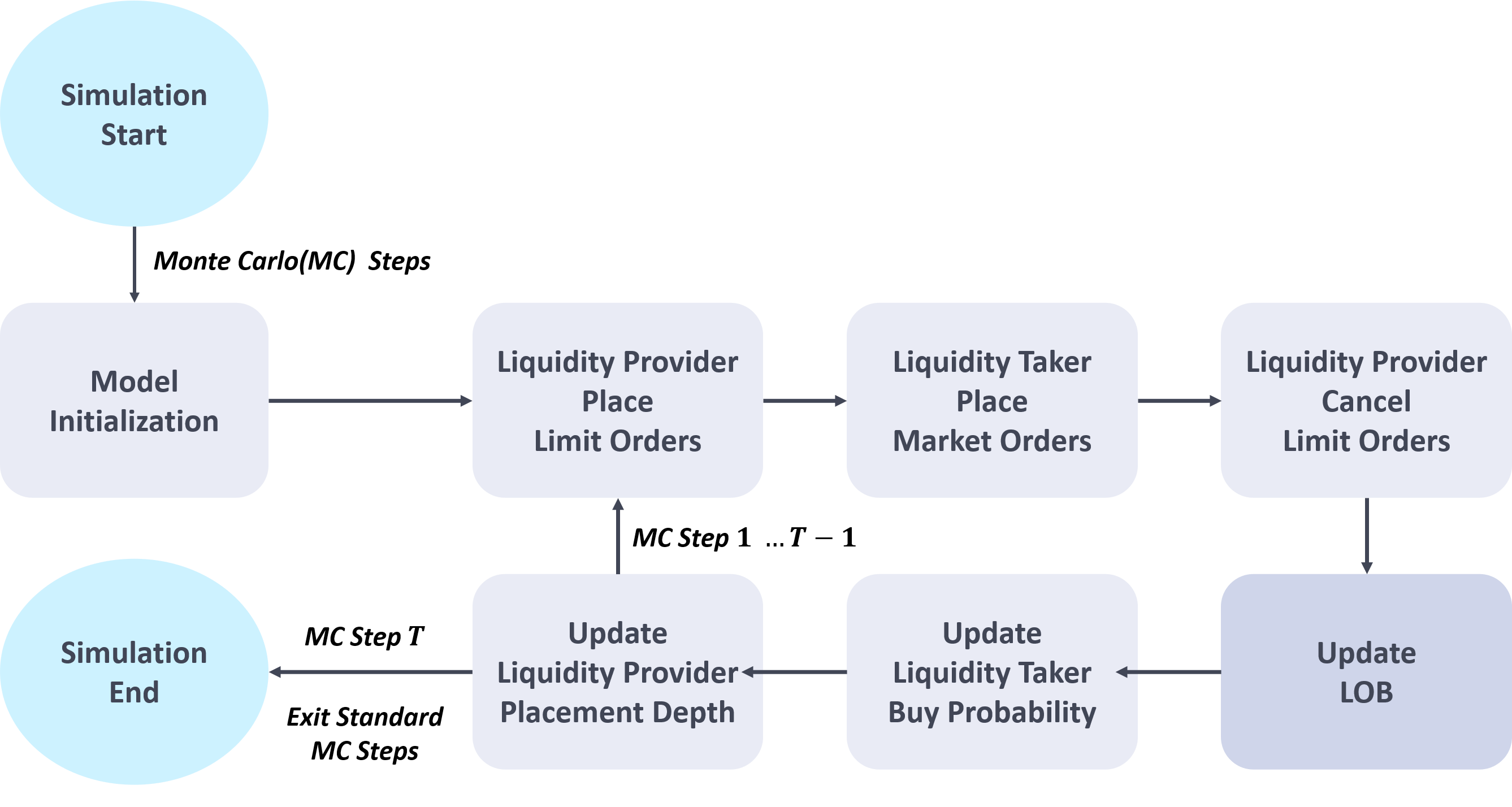}
\caption{The workflow of the PGPS financial market simulation model.}
\label{fig:PGPS-simulation}
\end{figure*}

\section{Empirical Studies}

To comprehensively evaluate the proposed SimLOB, several research questions (RQs) naturally arise:\\
\textbf{RQ1}: How well does SimLOB represent the latent structure of the underlying market dynamics?\\
\textbf{RQ2}: Can SimLOB generalize across different financial markets?\\
\textbf{RQ3}: Are the learned latent representation interpretable for financial applications?\\
\textbf{RQ4}:  What is the recommended architecture?

To address these questions, we design four groups of experiments:\\
\textbf{Group 1}: Assessing the representation capability of SimLOB on synthetic training data, focusing on reconstruction and calibration tasks.\\
\textbf{Group 2}: Evaluating the generalization of SimLOB on real stock data and synthetic data with advanced order types and varying tick sizes, while trained on merely synthetic data.\\
\textbf{Group 3}: Investigating the interpretability of different network layers of SimLOB.\\
\textbf{Group 4}: Ablating the parameter sensitivity and structural design.

\subsection{Group 1: Representation on Synthetic Data for Calibration}

We conduct three main experiments in \textbf{Group 1}. First, the proposed SimLOB and eight state-of-the-art networks are trained on a large volume of synthetic LOB data generated by an FMS model, as introduced in Section~\ref{sec:general-setup}. The reconstruction errors on the test set are then measured to evaluate the effectiveness of the vectorized representations. Second, we show that calibrating FMSs with learned representations of LOB is more beneficial than raw LOB inputs and traditional mid-price signals. Third, latent vectors from SimLOB and 8 compared networks are applied to ten FMS calibration tasks to examine whether better LOB representations consistently lead to improved calibration performance.











\begin{table}[b]
\centering
\caption{Parameter Ranges for Generating Synthetic Data}
\begin{tabular}{c||c|p{4.7cm}} \hline\hline
\textbf{Para.} & \textbf{Range} & \textbf{Remarks} \\\hline\hline
$\lambda_0$ &$[1,200]$ & \scriptsize{Controls price of each limit order}.\\
$C_{\lambda}$ &$[1,20]$ & \scriptsize{Controls price of each limit order}.\\
$\Delta_S$  &$[0.0005,0.003]$ & \scriptsize{Controls side probability of market orders.}\\
$\alpha$ &$[0.05,0.45]$ & \scriptsize{Probability liquidity providers submit orders.}\\
$\mu$ &$[0.005,0.085]$ & \scriptsize{Probability liquidity takers submit orders}.\\
$\delta$ &$[0.005,0.05]$ & \scriptsize{Probability liquidity takers cancel orders}.\\
\hline\hline 
\end{tabular}
\label{table:The ranges for randomly sampling parameters}
\end{table}

\subsubsection{General Experimental Setup}
\label{sec:general-setup}
\textbf{The Adopted FMS Model.} The widely studied Preis-Golke-Paul-Schneid (PGPS) model \cite{preis2006multi,yang2025towards} is adopted as the FMS model, simulating two types of agents: 125 liquidity providers submiting limit orders and  125 liquidity takers submiting market orders or cancel untraded limit orders. 
The simulation flowchart is shown in Fig. \ref{fig:PGPS-simulation}. 
At each time step $t$-th, a liquidity provider submits a limit order with probability $\alpha=0.5$.
A liquidity taker submits a market order with probability $\mu$ and cancel an untraded limit order with probability $\delta$. 
The probability of the market order being either the bid side or ask side is $q_{taker}(t)$ and $1-q_{taker}(t)$, respectively, where $q_{taker}(t)$ follows a mean-reverting random walk centered at 0.5, with increments of $\pm \Delta s$ and reversion probability $0.5 + |q_{taker}(t) - 0.5|$
Each order has a fixed volume of 100 shares.
The price of an ask limit order is determined by $p^a_1(t) - \lfloor -\lambda(t) \log u)\rfloor - 1$ and a bid limit order is determined by $p^b_1(t) - \lfloor -\lambda(t) \log u)\rfloor + 1$, where $\lambda(t) = \lambda_0(1 + \frac{|q_{taker}(t) - 0.5|}{\sqrt{\langle q_{taker} - 0.5\rangle^2}}C_{\lambda})$.
Here $\langle q_{taker} - 0.5\rangle^2 $ indicates a pre-computed value obtained by taking the average of $10^5$ Monte Carlo iterations of $(q_{taker} - 0.5)^2$ before simulation. 
$u \sim U(0,1)$ is a uniform random number.
The price of the market order is automatically set to the best price level on the opposite side.
In summary, the PGPS model contains 250 agents and the key parameters to be calibrated are $\mathbf{w}=[\delta, \lambda_0, C_{\lambda}, \Delta_s, \alpha, \mu]$.

\textbf{The Dataset for Representation Learning.} Augmenting training data with the synthetic LOB generated by FMS is increasingly popular \cite{assefa2020generating,cao2022synthetic}.
Hence, this work trains the SimLOB with purely the synthetic LOB data generated by the PGPS model with different parameters. 
In general, each setting of the 6-parameter tuple of PGPS actually defines a specific market scenario.
To ensure the synthetic data enjoys enough diversity, we uniformly randomly sample 2000 different settings of those 6-parameter tuples from a predefined range suggested by \cite{platt2018can} (see Table \ref{table:The ranges for randomly sampling parameters}). 
Each configuration simulates over 50,000 time steps, yielding 500 sequences with 100 time steps. 
This results in 1 million LOB sequences ($\tau=100$), with 80\% for training and  20\% for testing.
Additionally, real market data from the sz.000001 stock of the Chinese market (during May of 2019) is used for evaluation.

\begin{table}[t]
\setlength{\tabcolsep}{2pt} 
\centering
\caption{The 9 Compared Networks}
\resizebox{\columnwidth}{!}{ 
\begin{tabular}{r||cccp{4.0cm}} \hline\hline
 & Time series's length $\tau$ & The size of $\mathbf{x}(t)$ &$\#$trainable parameters \\\hline\hline
MLP & 100 & 40 & 2.2E7 \\
LSTM & 100 & 40 & 1.6E4 \\
CNN1 & 100 & 40 & 3.5E4 \\
CNN2 & 300 & 40 & 2.8E5 \\
CNN-LSTM & 300 & 42 & 5.3E4 \\
DeepLOB & 100 & 40 & 1.4E5 \\
TransLOB & 100 & 40 & 1.1E5 \\
TransLOB-L & 100 & 40 & 1.5E7 \\
SimLOB & 100 & 40 & 1.3E7 \\ 
\hline \hline
\end{tabular}
}
\label{table:characteristics of the 9 compared networks}
\end{table}

\textbf{The Compared Networks.} 
This experiment compares 7 SOTA networks and one enhanced version (TransLOB-L), using their architecture (excluding the last linear layer) as representation learning layers, all configured with $\tilde{\tau}=128$. 
These networks have been frequently adopted in comparisons in recent deep learning based LOB analysis works \cite{prata2024lob,huang2021benchmark,zhang2019deeplob}, primarily focusing on price trend prediction.
We follow the conventions of \cite{prata2024lob} to name the 8 networks as MLP\cite{tsantekidis2017using}, LSTM\cite{tsantekidis2017using}, CNN1\cite{nino2019cnn}, CNN2\cite{tsantekidis2020using}, CNN-LSTM\cite{tsantekidis2020using}, DeepLOB\cite{zhang2019deeplob}, TransLOB\cite{wallbridge2020transformers}. 
As their names suggest, despite the fact that MLP, LSTM, CNN1, and CNN2 adopt single-type networks, the other works mostly employ a Convolution-Recurrent architecture to intuitively first extract the prices precedence and then capture the temporal auto-correlation.
TransLOB employs the Transformer architecture to replaces the recurrent network while still keeping the convolution layers.
TransLOB-L is an enlarged version of TransLOB to keep the size approximately aligned with SimLOB by using 7 stacked Transformers instead of 2 as in the original TransLOB.
All 8 networks are trained under the same protocol as SimLOB, with
input formats and sizes follow the original papers. Table \ref{table:characteristics of the 9 compared networks} summarizes their characteristics:
the second and third columns list acceptable input formats of each network, and the last column shows the sizes of the networks.
The MLP has the largest size of $2.2 \times 10^7$ weights, which can be simply considered as the enlarged version of $\mathrm{FCN}_2$.


\textbf{The Calibration Tasks.} Note that $M(\mathbf{w})$ is basically a software simulator.
In this work, it is programmed on the Multi-Agent Exchange Environment (MAXE)\footnote{https://github.com/maxe-team/maxe} environment developed by the University of Oxford with the MIT License \cite{belcak2021fast}.
Thus the calibration problem of $\min_{\mathbf{w}} D(\hat{\mathbf{X}}_T, M(\mathbf{w}))$ lacks mathematical properties like gradients, making black-box optimization suitable.
For simplicity, a standard Particle Swarm Optimizer (PSO)\cite{shi1998modified} is employed with recommended hyper-parameters: a population size is set to 40, the inertia weight is set to 0.8, and the cognitive and social crossover parameters $c1 = 0.5, c2 = 0.5$.
The total iteration number of PSO in each run is fixed to 100 \cite{platt2020comparison}.

\begin{table}[t]
\centering
\caption{Parameter Settings of Generating 10 Synthetic Data with the PGPS Model}
    \begin{tabular}{r||rrrrrr}\hline\hline
    \textbf{Paras}  & \textit{$\mathbf{\lambda_0}$} &  \textit{$\mathbf{C_{\lambda}}$} & \textit{$\mathbf{\alpha}$} &  \textit{$\mathbf{\mu}$} &  \textit{$\mathbf{\Delta_S}$} & \textit{$\mathbf{\delta}$}  \\ \hline\hline
        data 1 & 80 & 8 & 0.1 & 0.02 & 0.002 & 0.02\\
        data 2 & 120 & 11 & 0.2 & 0.03 & 0.003 & 0.03\\
        data 3 & 130 & 12 & 0.3 & 0.04 & 0.003 & 0.04\\
        data 4 & 90 & 9 & 0.15 & 0.02 & 0.001 & 0.02\\
        data 5 & 70 & 7 & 0.15 & 0.015 & 0.0015 & 0.03\\
        data 6 & 134.64 & 15.45 & 0.3275 & 0.07116 & 0.002 & 0.0324\\
        data 7 & 17.7 & 11.36 & 0.2639 & 0.067 & 0.00066 & 0.03644\\
        data 8 & 153.53 & 9.27 & 0.2983 & 0.07343 & 0.00238 & 0.01278\\
        data 9 & 48.13 & 3.54 & 0.4374 & 0.02645 & 0.00077 & 0.01674\\
        data 10 & 7.13 & 7.04 & 0.1106 & 0.05609 & 0.00217 & 0.01389\\
        \hline\hline
\end{tabular}
\label{table:Parameter settings of generating 10 synthetic data with the PGPS model}
\end{table}

10 synthetic LOB data are randomly generated with 10 different settings of 6-parameter tuples, each containing $T=3600$ time steps to simulate approximately 1 hour of market activity. The parameters range for randomly sampling parameters is in the Table \ref{table:The ranges for randomly sampling parameters}, and the sampled tuples are listed in Table \ref{table:Parameter settings of generating 10 synthetic data with the PGPS model}. These data are utilized as the target data.
Note that these data instances are challenging that they have a much higher frequency than traditional FMS works who can only calibrate to daily data \cite{platt2018can}.
And the length of the target data is also at least $10\times$ longer than the existing calibration works \cite{platt2020comparison}.
The objective function for calibration employs MSE. 
The calibration objective for mid-prices follows traditional works do, Eq.(\ref{eq: traditional calibration objective}) gives

\begin{equation}
D(\hat{\mathbf{X}}_T, M(\mathbf{w})) = \frac{S_1(\mathbf{w})}{\lceil \frac{T}{\tau} \rceil},
\label{eq: traditional calibration objective}
\end{equation}
\begin{equation*}
S_1(\mathbf{w}) = \sum^{\lceil \frac{T}{\tau} \rceil}_{i=1}\sum^{\tau}_{t=1} (\hat{mp}((i-1)\tau+t) - mp^\mathbf{w}((i-1)\tau+t))^2,
\end{equation*}

\noindent where $M(\mathbf{w})=\mathbf{X}_t=\{\mathbf{x}(t)\}^T_{t=1}$. 
For calibrating latent vectors, this paper adopts:

\begin{equation}
D(\hat{\mathbf{X}}_T, M(\mathbf{w})) = \frac{S_2(\mathbf{w})}{\lceil \frac{T}{\tau} \rceil},
\label{eq: calibration with latent vector}
\end{equation}

\begin{equation*}
S_2(\mathbf{w}) = \sum^{\lceil \frac{T}{\tau} \rceil}_{i=1} \sum^{\tilde{\tau}}_{t=1}(\hat{z}((i-1)\tilde{\tau}+t) - z^\mathbf{w}((i-1)\tilde{\tau}+t))^2,
\end{equation*}

\noindent where $\tau=100$ and $\tilde{\tau}=128$.

\begin{figure}[t]
\centering
\includegraphics[width=3.5in]{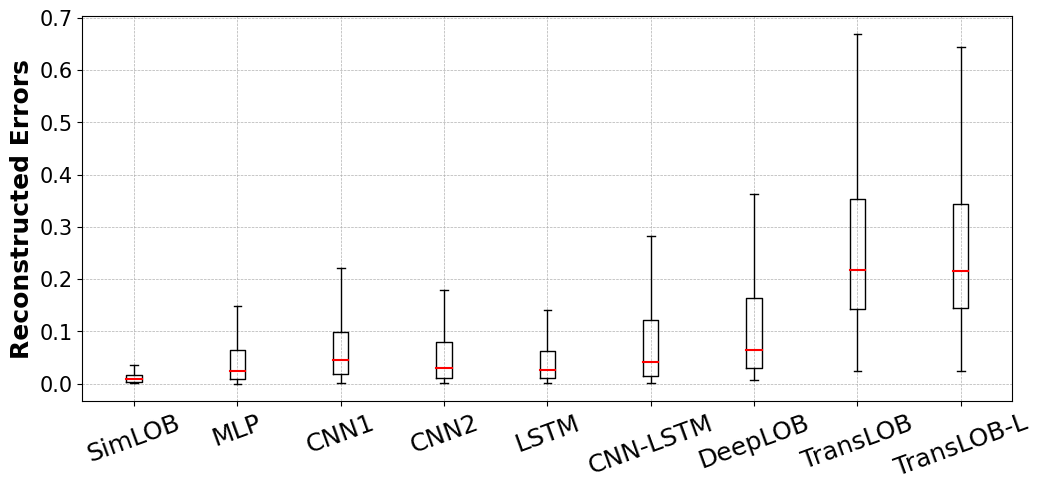}
\caption{Reconstruction errors on 0.2M testing instances.}
\label{fig:boxplot of reconstruction error}
\end{figure}

\textbf{Computational Resources.} The experiments run on a server with 250GB memory, an Intel(R) Xeon(R) Gold 5218R CPU @ 2.10GHz with 20 physical cores, and 2 NVIDIA RTX A6000 GPUs. 
Training of each network on average required approximately 50 hours using data parallelism on the 2 GPUs.
Calibrating each LOB sequence costs 2-3 hours on 20 CPU cores (40 threads).
The simulator runs on a single core, while the PSO can be run in parallel with 40 simulators.

\begin{figure*}[t]
    \centering
    \includegraphics[width=\textwidth, height=0.45\textwidth]{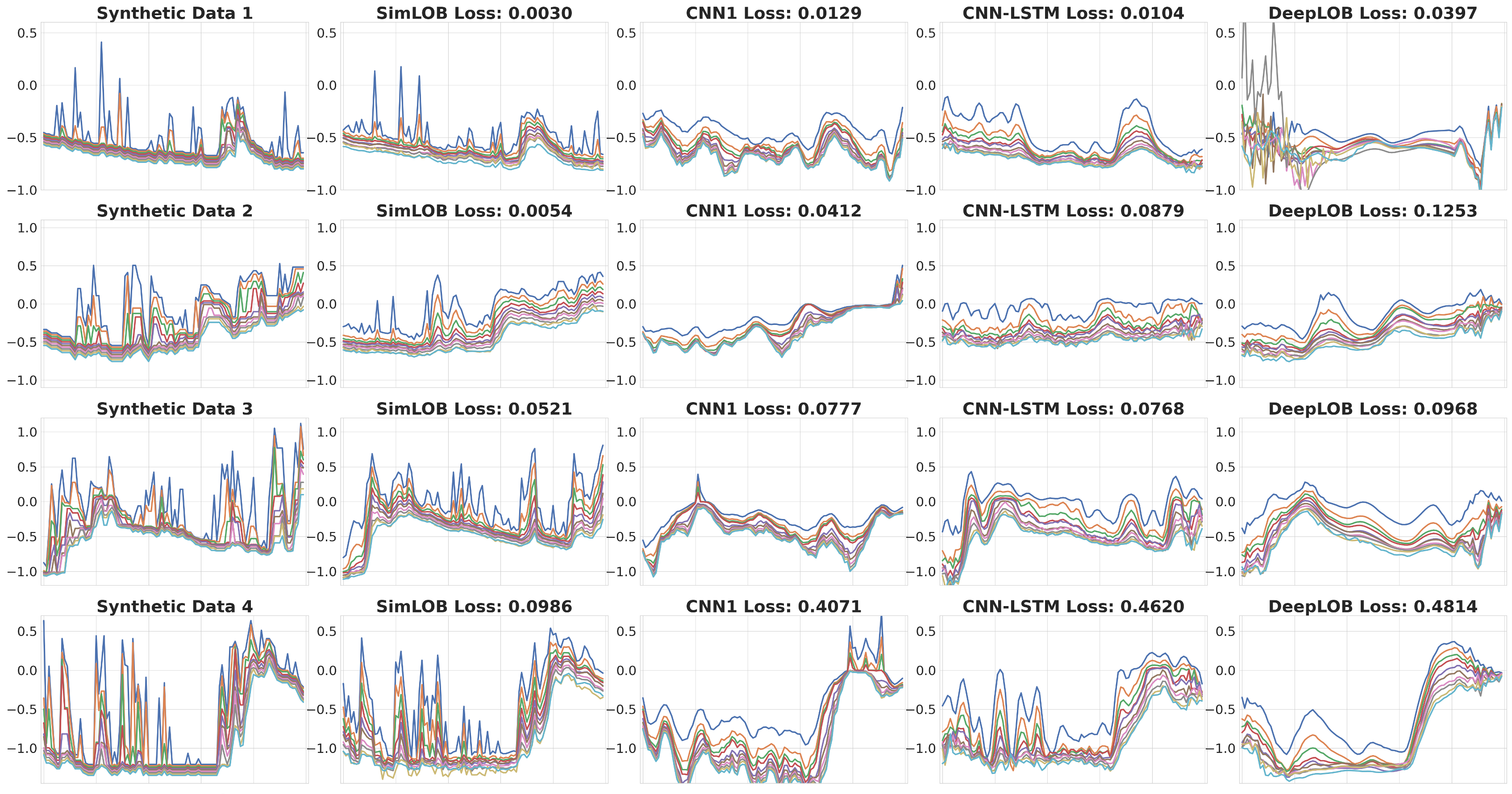}
    \caption{{The reconstructed bid prices (colored lines represent 10 levels) on 4 synthetic data are visualized with $\mathbf{Err}_r$.}}
    \label{fig:reconsturct bid price1}
\end{figure*}



\subsubsection{Experiment-1: Learning the Vectorized Representations of LOB}
The reconstruction errors on 0.2 million synthetic testing data of all 9 networks are depicted in Fig. \ref{fig:boxplot of reconstruction error}.
Among them, SimLOB not only achieves the smallest averaged reconstruction error, but performs the most stably, with reconstruction errors of SimLOB following a power-law distribution and a mode of 0.003.
Almost all the CNN-based networks (including CNN1, CNN2, CNN-LSTM, TransLOB, and TransLOB-L) perform the worst, suggesting that CNN is not effective as expected to deal with LOB data.
This is quite contradictory to the trend of existing works on adopting deep learning for analyzing LOB.
This phenomenon is later studied in-depth in Section IV-E with more empirical ablations.

The reconstructed data of seven compared networks across four synthetic data demonstrates the superiority of SimLOB.
These depicted data is randomly selected from the testing instances based on the principles of covering different fluctuating amplitudes and trends.The data reconstructed by SimLOB mostly closely resembles the ground truth, while other networks have tended to smooth the target data, failing to fully preserve the key properties of LOB in their latent vectors. TransLOB and TransLOB-L are excluded from visualization as their reconstructed data appear to be nearly flat, consistent with their poor performance shown in Fig. \ref{fig:boxplot of reconstruction error}.
This indicates that the Transformer architecture alone is not the panacea to the representation learning of LOB.
And thus it verifies that the importance of the proposed feature extraction and dimension reduction blocks in SimLOB.



\begin{figure}[thb]
\centering
\includegraphics[width=3.5in,height=2in]{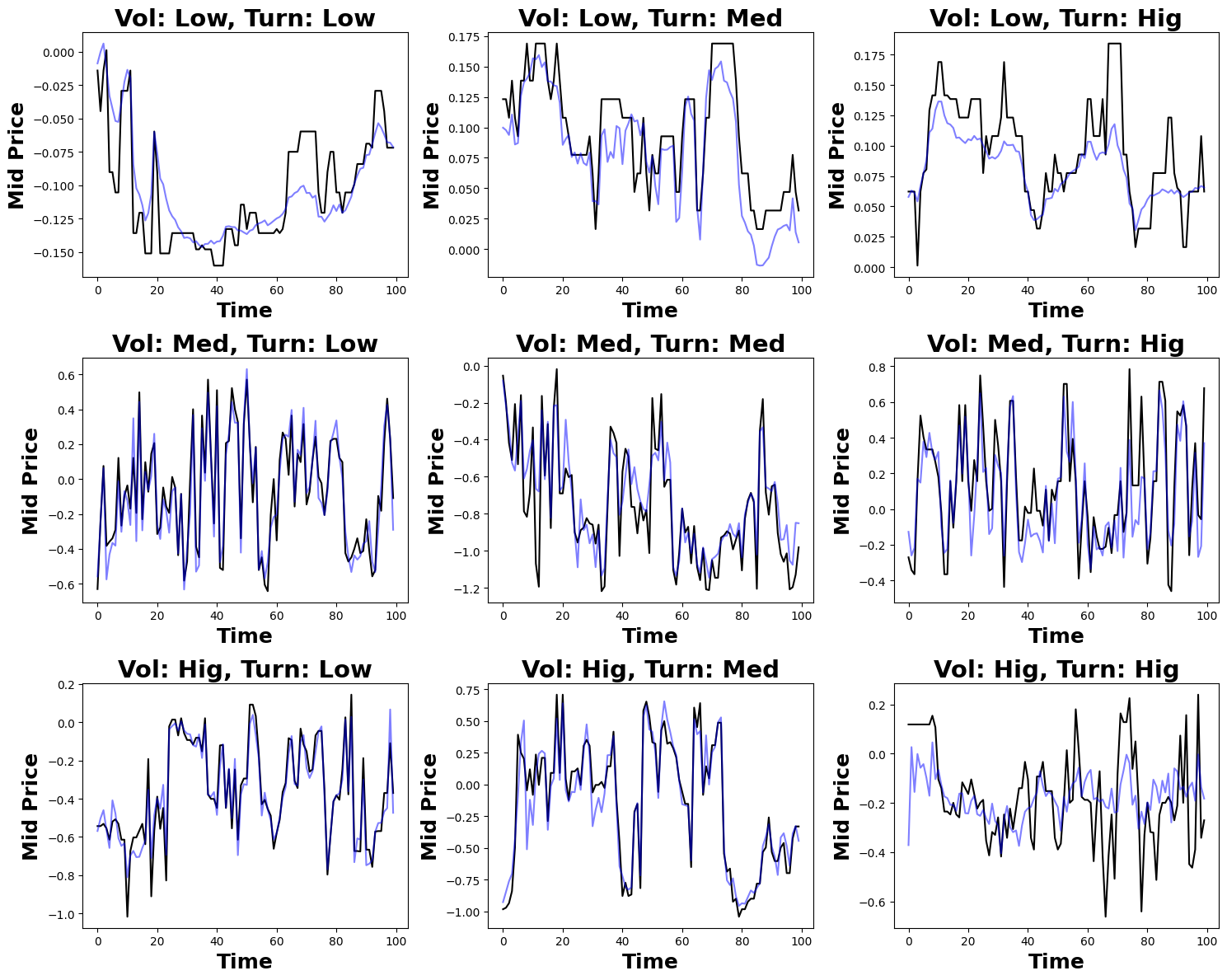}
\caption{{Mid-price reconstruction on synthetic data with different levels of volatility and turnover.}}
\label{fig:reconstruc_synthetic}
\end{figure}

\begin{figure*}[t]
    \centering
    \includegraphics[width=\textwidth, height=0.75\textwidth]{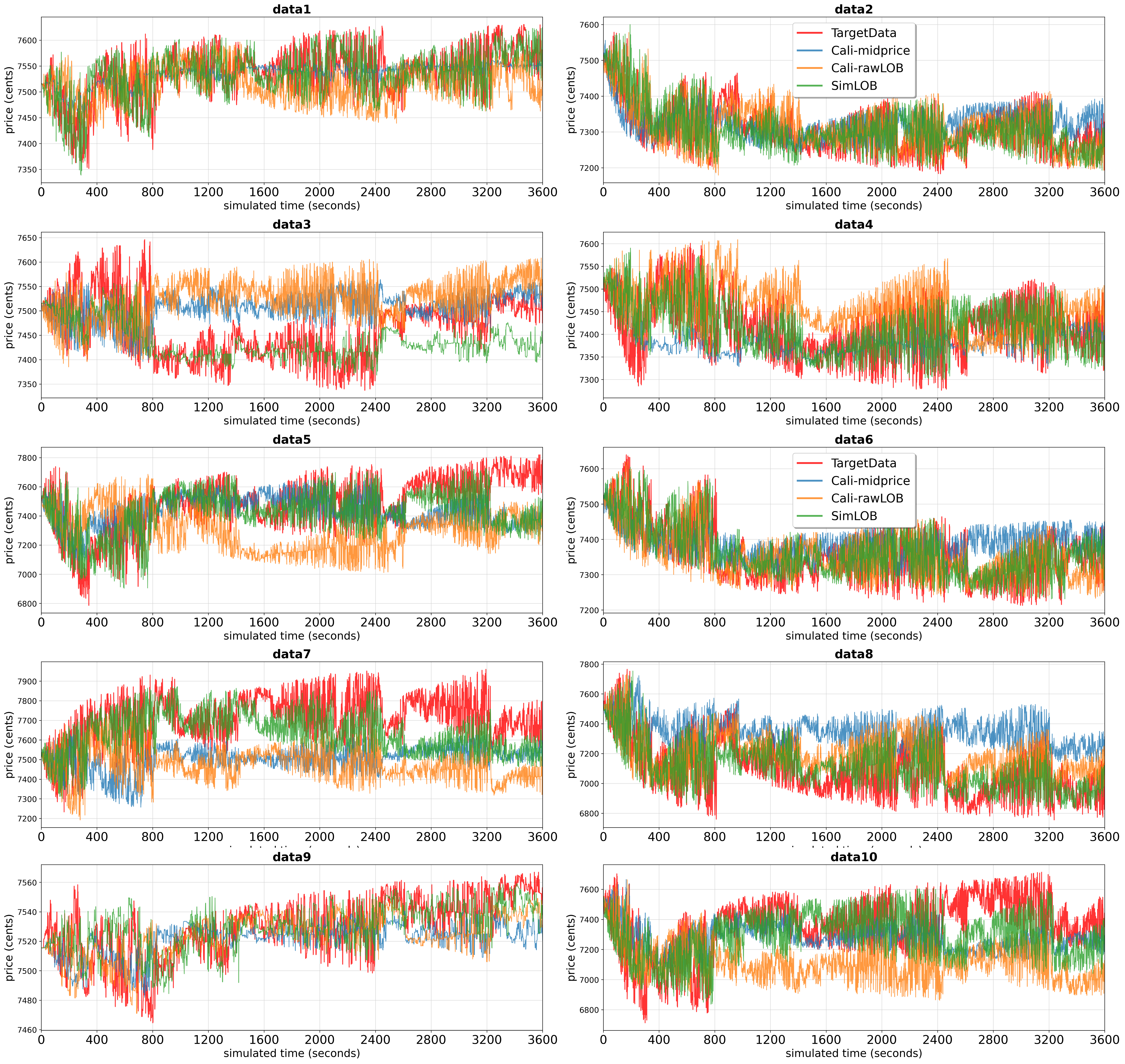}
    \caption{Mid-prices of the \textcolor{red}{Target data}, \textcolor{green}{Cali-SimLOB}, \textcolor{orange}{Cali-rawLOB}, and \textcolor{blue}{Cali-midprice} on 10 calibration instances. }
    \label{fig:midprice curves}
\end{figure*}

To show the robustness of SimLOB on various market conditions, we conducted reconstruction experiments using synthetic data with varying volatility and turnover levels. We categorized the synthetic data into three severity levels: low, medium, and high, based on its volatility and turnover values.
As illustrated in Fig. \ref{fig:reconstruc_synthetic}, the reconstruction results under different combinations of volatility and turnover. The black lines represent the ground truth mid-price series, while the blue lines show the reconstructed mid-price. It can be seen that SimLOB consistently provides accurte reconstructions across all combinations of volatility and turnover. This indicates that SimLOB effectively captures the mid-price dynamics, demonstrating robustness and adaptability to varying market conditions.

\begin{table}[b]
\centering
\caption{Calibration Errors using Three Types of LOB Representations on 10 Synthetic Data}
\begin{tabular}{l||p{1.65cm} p{1.65cm} p{1.65cm}} \hline\hline
       & Cali-SimLOB            & Cali-rawLOB       & Cali-midprice \\\hline\hline
data1  & \textbf{1.29E03} & 1.01E04          & 7.84E03  \\
data2  & 2.70E03          & \textbf{2.26E03} & 1.29E04  \\
data3  & \textbf{4.35E02} & 5.56E02          & 3.04E03  \\
data4  & \textbf{5.51E02} & 6.18E02          & 3.89E03  \\
data5  & \textbf{1.20E03} & 6.12E03          & 5.07E03  \\
data6  & \textbf{8.58E03} & 3.29E04          & 2.71E04  \\
data7  & \textbf{1.49E06} & 1.81E06          & 2.01E06  \\
data8  & \textbf{1.07E03} & 2.48E03          & 2.54E03  \\
data9  & \textbf{1.13E04} & 3.79E04          & 1.68E04  \\
data10 & \textbf{9.83E02} & 1.49E03          & 2.88E03 \\\hline\hline
\end{tabular}
\label{table:Calibration with SimLOB, Cali-rawLOB, midprice}
\end{table}

\subsubsection{Experiment-2: Calibrating Vectorized LOB is Beneficial}
To demonstrate that calibrating LOB is beneficial to FMS, the 10 synthetic data are also calibrated using traditional objective function Eq.(\ref{eq: calibration with latent vector}) with merely mid-prices, denoted as Cali-midprice.
Furthermore, one may concern why not directly use the reconstruction error $\mathbf{Err}_r$ as the calibration function to calibrate the raw LOB.
Hence, we also calibrate PGPS to the raw LOB, denoted as Cali-rawLOB. 

Table \ref{table:Calibration with SimLOB, Cali-rawLOB, midprice} shows the calibration with the representations learned by SimLOB (denoted as Cali-SimLOB) outperforms Cali-midprice and Cali-rawLOB in 9 out of 10 instances.
For data 2, SimLOB achieves competitive results compared to Cali-rawLOB and is significantly better than Cali-midprice.
These results confirm that embedding LOB is beneficial to the calibration of FMS.
In contrast, no clear advantage between Cali-rawLOB and Cali-midprice, possibly explaining why traditional FMS studies focused only on mid-price. Mid-prices are simpler to handle and yield similar calibration results with raw LOB data.

The simulated mid-prices of Cali-SimLOB, Cali-rawLOB, and Cali-midprice on 10 target data are also depicted in Fig. \ref{fig:midprice curves}.
It can be intuitively seen that the simulated mid-price of SimLOB resembles the target much better than that of Cali-rawLOB and Cali-midprice.
Comparing Cali-SimLOB with Cali-rawLOB, $\mathbf{Err}_r$ measures the MSE between two long vectors ($40\tau=4000$) for each segmentation.
Though it captures differences in raw LOB, it does not reflect the importance of the key properties of temporal auto-correlation and price precedence.
Compared to Cali-midprice, SimLOB helps PGPS achieve much better performance on not only the whole LOB but also the mid-price, simply because the mid-price is basically a derivative from LOB.
To summarize, calibrating with more information of the market improves the fidelity of the simulation, but needs effective latent representations.
This supports the initial motivation of this work.

\begin{figure}[t]
\centering
\includegraphics[width=2.8in]{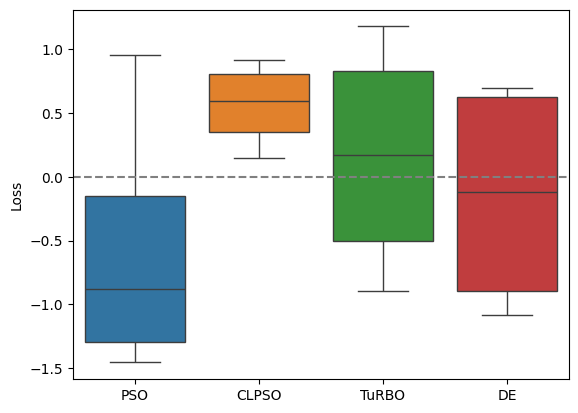}
\caption{{Calibration errors with different optimization algorithms.}}
\label{fig:algo opt}
\end{figure}

We also compared typical optimization algorithms, including PSO, comprehensive learning particle swarm optimizer (CLPSO) \cite{clpso}, differential evolution (DE) \cite{de} and  trust region bayesian optimization(TurBO) \cite{eriksson2019scalable}. This aims to investigate the potential impact of the optimization algorithm on the overall performance.
For consistency, we calibrated all four algorithms on the same four synthetic datasets, ensuring identical computational budgets by keeping the number of evaluations constant across methods. The errors were then normalized using Z-score to account for scale differences among tasks and visualized using box plots in Fig. \ref{fig:algo opt}, which demonstrate that PSO consistently achieves competitive performance compared to the other algorithms. 

\begin{table*}[b]
\centering
\renewcommand{\arraystretch}{1.1} 
\caption{Calibration Errors of 8 Networks on 10 Synthetic Data}\label{table:Comparisons among the 9 Compared Networks on 10 Synthetic Data}
\normalsize
\begin{tabular}{l||*{5}{p{1.3cm}}*{1}{p{1.5cm}}*{2}{p{1.45cm}}*{3}{p{1.3cm}}}\hline\hline
       & SimLOB           & MLP              & CNN1       & CNN2        & LSTM          & CNNLSTM & DeepLOB  & TransLOB \\ \hline\hline
data1  & 1.27E03          & 2.41E03          & 1.69E03     &1.65E03     &\textbf{8.98E02} & 2.32E03  & 9.60E03  & 3.53E03 \\
data2  & \textbf{1.30E03} & 1.85E03          & 1.61E03     &5.67E03     &2.43E04         & 1.51E04  & 4.14E03  & 2.52E04 \\
data3  & \textbf{4.35E02} & 6.95E02          & 6.05E02     &6.57E03     &1.13E03         & 1.47E03  & 2.18E03  & 2.55E03 \\
data4  & \textbf{5.51E02} & 1.10E03          & 2.79E03     &1.53E03     &4.57E03         & 8.27E02  & 2.27E03  & 1.15E04 \\
data5  & \textbf{1.20E03} & 8.70E03          & 1.19E04     &1.74E03     &2.49E04         & 4.72E03  & 1.63E04  & 4.77E03 \\
data6  & \textbf{8.58E03} & 2.11E04          & 2.87E04     &1.34E04     &1.07E05         & 3.40E04  & 8.42E04  & 3.29E04 \\
data7  & 1.49E06          & \textbf{3.07E05} & 1.11E06     &1.09E06     &3.74E06         & 1.62E06  & 1.92E06  & 1.60E06 \\
data8  & \textbf{1.07E03} & 2.38E03          & 3.03E03     &1.84E03     &1.37E04         & 2.61E03  & 8.80E03  & 5.08E03 \\
data9  & 1.13E04          & 1.97E04          & \textbf{8.42E03} &1.46E04 &6.68E04         & 2.21E04  & 6.41E04  & 2.68E04 \\
data10 & \textbf{9.83E02} & 1.36E03          & 1.98E03     &4.30E04     &1.31E04         & 8.28E03  & 1.51E04  & 4.12E03 \\ \hline\hline
\end{tabular}
\end{table*}

\begin{figure}[t]
\centering
\includegraphics[width=3.5in]{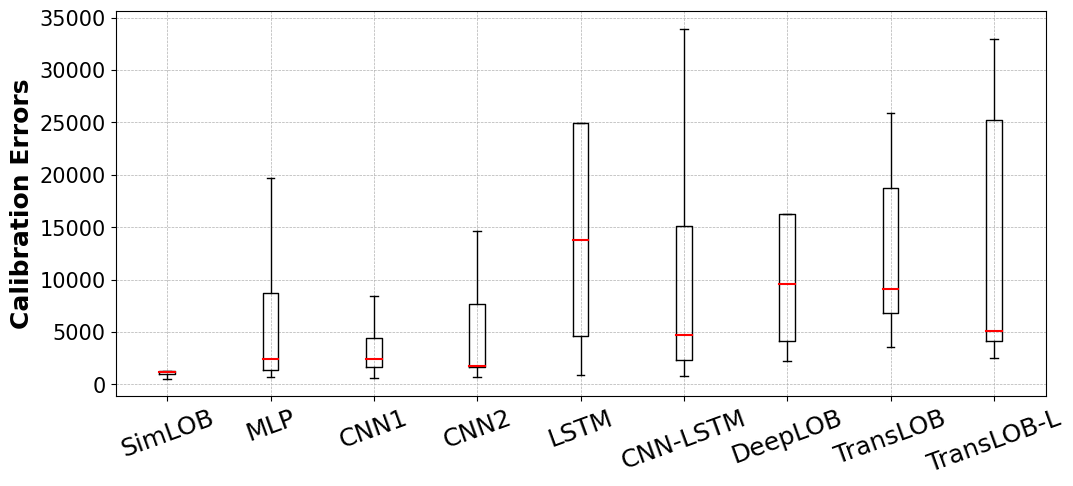}
\caption{Calibration errors on testing instances.}
\label{fig:boxplot of calibration error}
\end{figure}

\subsubsection{Experiment-3: Better Representation, Better Calibration}
Each of the 10 synthetic data ($T=3600$) is processed by 9 networks to obtain the latent vectors.
Using Eq.(\ref{eq: calibration with latent vector}) and PSO, the PGPS is calibrated to find the optimal parameter tuples and the corresponding simulated data.
Calibration errors ($\mathbf{Err}_r$) between the target and the simulated data are presented in Table \ref{table:Comparisons among the 9 Compared Networks on 10 Synthetic Data}.
Among all representations, SimLOB helps achieve the best calibration performance on 7 out of 10 instances and the runner-up performance on data 1 and data 9.
The distribution of calibration errors (outliers omitted) in Fig. \ref{fig:boxplot of calibration error} shows the advantages of SimLOB over the compared networks. 
Furthermore, Fig. \ref{fig:boxplot of reconstruction error} and Fig. \ref{fig:boxplot of calibration error} jointly reveal a consistency between the reconstruction errors and calibration errors that better representations more likely lead to better FMS calibration.

\subsection{Group 2 – Generalization of SimLOB to Unseen Data}

Due to the non-stationary nature of financial market, it is important that SimLOB can generalize to unseen data. 
In this group, we still train the networks on previously mentioned synthetic data. Then we use the encoder of each network to embed three types of unseen data: 1) real LOB data from NASDAQ (USA) and Shenzhen (China) stock exchanges, 2) synthetic LOB data generated by another FMS model and injected with advanced order types such as \textbf{ICEBERG}, \textbf{TWAP}, and \textbf{VWAP}, and different tick sizes. After that, the decoder of each trained network is used to reconstruct LOB from the embeded latent. 
The reconstruction errors ($\mathbf{Err}_r$) between the original LOB and the reconstructed LOB is measured.

\subsubsection{Reconstruction on Real Stock Data}
\label{sec:Reconstruction on NASDAQ}
Four representative NASDAQ stocks are selected: \textbf{Cisco Systems (CSCO)}, \textbf{Intel Corporation (INTC)}, \textbf{Booking Holdings (PCLN)}, and \textbf{Tesla (TSLA)}. For each stock, we collected data from January 1 to March 31 in 2015. To show the representativeness and diversity of the collected data, we calculated daily volatility and turnover of each stock, and then divided the dates into different market conditions: low, medium, and high levels of volatility and turnover, respectively.

\begin{figure}[!t]
\centering
\includegraphics[width=3.5in,height=2.6in]{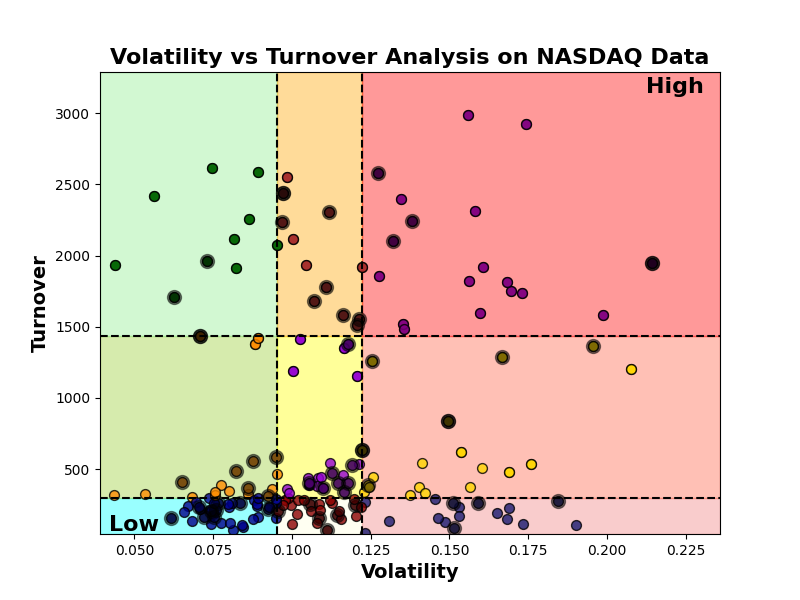}
\caption{Scatter plot of NASDAQ stocks categorized by volatility and turnover. }
\label{fig:NASDAQ different regimes}
\end{figure}

\begin{table*}[htbp]
\centering
\caption{Reconstruction Errors on NASDAQ data}
\label{tab:Reconstruction error on NASDAQ different regimes}
\resizebox{\textwidth}{!}{
\begin{tabular}{llccccccc}
\toprule
\multirow{2}{*}{\textbf{Market Conditions}} & \multirow{2}{*}{\textbf{Severity}} & \multicolumn{7}{c}{\textbf{Networks}} \\
\cmidrule(lr){3-9}
& & SimLOB2 & SimLOB8 & MLP & CNN1 & CNN2 & TransLOB & DeepLOB \\
\midrule

\multirow{4}{*}{\textbf{Volatility}} 
& High & \textbf{0.0043} & 0.0072 & \underline{0.0048} & 0.0175 & 0.0072 & 0.0501 & 0.0211 \\
& Medium & \textbf{0.0060} & 0.0093 & \underline{0.0063} & 0.0201 & 0.0099 & 0.0514 & 0.0220\\
& Low & \textbf{0.0053} & 0.0075 & 0.0059 & 0.0193 & 0.0095 & 0.0396 & 0.0200 \\

\midrule

\multirow{3}{*}{\textbf{Turnover}} 
& High & \textbf{0.0047} & \underline{0.0056} & 0.0059 & 0.0180 & 0.0078 & 0.0579 & 0.0220 \\
& Medium & \underline{0.0061} & 0.0101 & \textbf{0.0054} & 0.0196 & 0.0104 & 0.0433 & 0.0208 \\
& Low & \textbf{0.0053} & 0.0075 & \underline{0.0059} & 0.0193 & 0.0095 & 0.0396 & 0.0200 \\

\bottomrule
\end{tabular}
}
\end{table*}

Then 25\% of the collected data (marked in black dots) across 9 different regions is randomly sampled for the reconstruction study and showed in Fig. \ref{fig:NASDAQ different regimes}.
From each selected date, one sequence of 15000 time steps of LOB snapshots were extracted to serve as the target data. The reconstruction errors of different networks on different market conditions are listed in Table \ref{tab:Reconstruction error on NASDAQ different regimes}, where the bold font indicates the best averaged error on each conditioned dataset, while the runner-up is underlined. To assess the robustness of our model, we consider two versions: SimLOB2, which uses two Transformer blocks, and SimLOB8, which adopts eight Transformer blocks.  



\begin{figure}[thb]
\centering
\includegraphics[width=3.5in,height=1.5in]{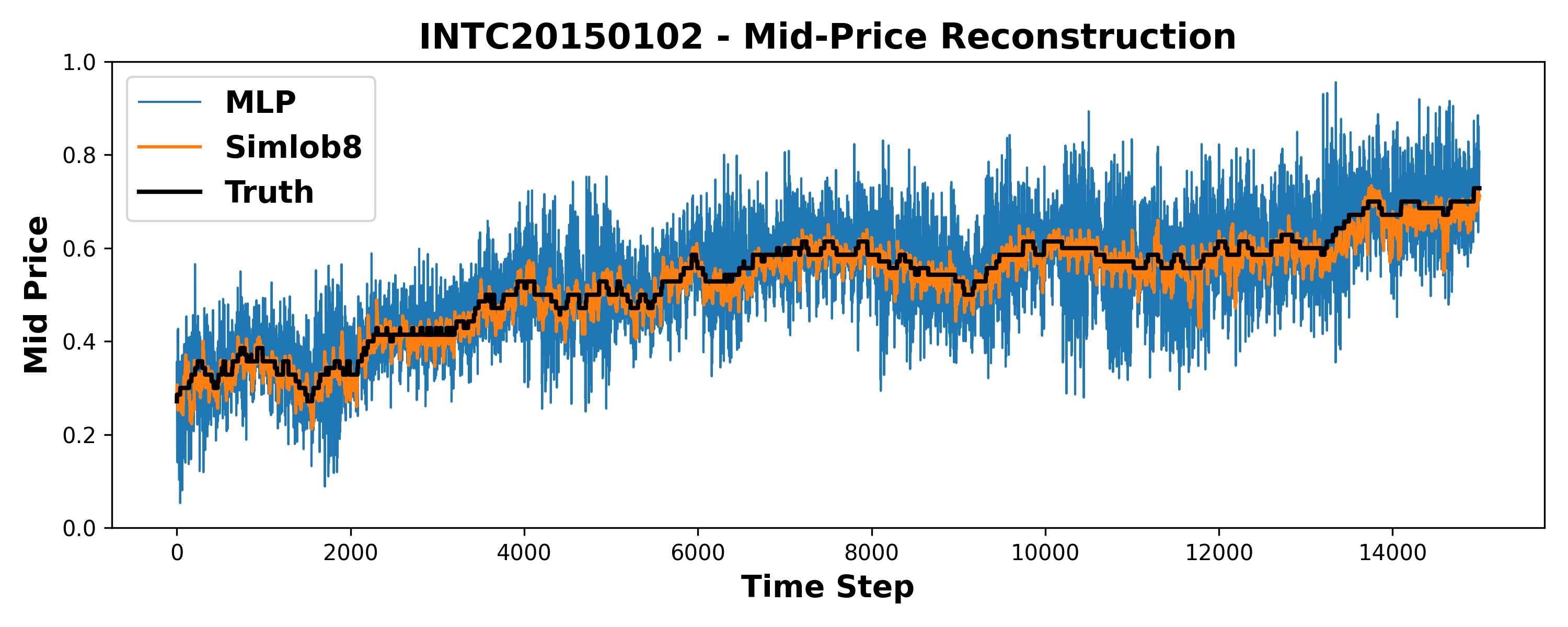}
\caption{
Reconstructed mid-price by SimLOB2 and MLP.}
\label{fig:SimLOB2 and MLP reconstruction curve on NASDAQ.}
\end{figure}

From Table \ref{tab:Reconstruction error on NASDAQ different regimes}, SimLOB2 achieves the lowest error in 3 out of 4 conditions, demonstrating its robust performance over the other networks. Comparatively, SimLOB8 performs poorer than SimLOB2 and even MLP on the reconstruction errors, while its reconstructed mid-price serie looks closer to the targe data (less volatile than that of MLP), as shown in Fig. \ref{fig:SimLOB2 and MLP reconstruction curve on NASDAQ.}. 
This observation suggests that more metrics can be adopted to distinguish the two networks in this kind of scenario. Hence, we additionally employ several statistical metrics commonly used in finance literature, i.e., stylized facts \cite{10.1145/3383455.3422561}: log return (LogR), volatility clustering (Vol. Clust.), volume-volatility correlation (Vol-Vol Corr), and autocorrelation (AutoCorr) on 16 random selected dates. We compare the Wasserstein distance between the log return distributions extracted from reconstructed and original mid-price series, as well as the differences in the absolute values of volatility clustering, volume-volatility correlation, and autocorrelation. The results are summarized in Table \ref{tab:reconstruction_stylized_fact}, where the rankings of the networks get changed. DeepLOB and TransLOB look much better than MLP, while SimLOB2 still demonstrates the best performance, indicating that SimLOB not only closely reconstructs the price series but also preserves statistical properties of the target data.

\begin{figure*}[!t]
\centering
\includegraphics[width=\textwidth, height=0.45\textwidth]{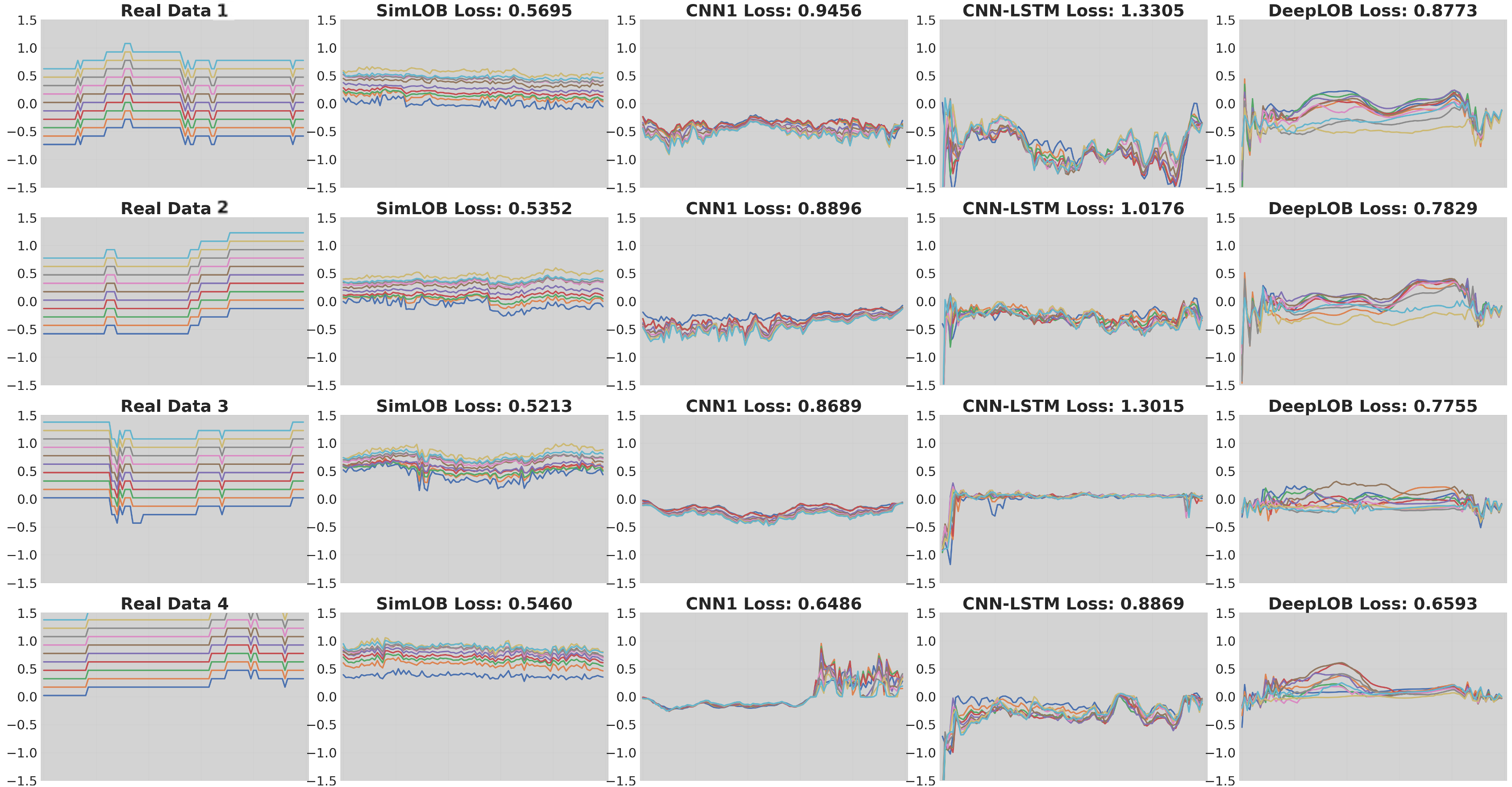}
\caption{
Reconstruction of the 10-level bid price series on PingAn Bank (SZ.000001).}
\label{fig:Reconsturc on SZ stock}
\end{figure*}

To evaluate the generalization ability of our model across different regional stock markets, we further present the reconstruction performance on SZ000001 (PingAn Bank), the first stock from the Shenzhen Stock Exchange. The 10 levels of bid prices in LOB are shown in Fig. \ref{fig:Reconsturc on SZ stock}, where each data represents 100 consecutive snapshots of LOB. It can be seen that SimLOB preserves the precedence constraints of the original 10 levels best.

\begin{table}[bht]
\centering
\caption{Comparison of Stylized Facts against the Ground Truth.}
\label{tab:reconstruction_stylized_fact}
\begin{tabular}{l||cccc}\hline\hline
Model & LogR (W) & Vol-Vol Corr & Vol. Clust. & AutoCorr \\\hline\hline
SimLOB2   & \textbf{0.000091} & \textbf{0.322103} & \textbf{0.114584} & 0.204052 \\
SimLOB8   & 0.000136          & 0.411049          & 0.187881           & 0.257303 \\
DeepLOB   & \underline{0.000098} & 0.376564 & \underline{0.174129} & \underline{0.196459} \\
CNN1      & 0.000478          & 0.422589          & 0.244086           & 0.487215 \\
CNN2      & 0.000164          & 0.384003          & 0.200089           & 0.271054 \\
MLP    & 0.000210          & 0.476237          & 0.225608           & 0.317023 \\
TransLOB  & 0.000200          & \underline{0.334227}          & 0.222856           & \textbf{0.154329} \\\hline\hline
\end{tabular}
\end{table}

\subsubsection{Reconstruction on Advanced Synthetic Stock Data}

In order to assess the reconstruction performance across different market configurations, we conduct experiments on synthetic data generated by a different FMS model as proposed by \cite{10.1145/3384441.3395986}. Specifically, the agents include one Exchange Agent, which maintains the LOB. Additionally, the configuration includes 100 Value Agents that observe a shared fundamental value time series with observation noise. They update their beliefs using Bayesian inference. Additionally, 25 Momentum Agents employ a simple momentum-based strategy: after every 50 observations, they compare the 50-step and 20-step moving averages (\( p_{50} \) and \( p_{20} \)), buying when \( p_{20} > p_{50} \) and selling otherwise. The simulation also incorporates 5000 Noise Agents, each of which places a single random buy or sell order during the trading day, with arrival time concentrated around market open and close to mimic real-world patterns. To ensure liquidity, a Percentage-of-Volume Market Maker Agent is also included, which adaptively maintains a no-trade window around the mid-price and posts limit orders outside this window, with order sizes proportional to the observed transaction volume over a 10-second lookback window.
Hences, we inject three types of advanced orders as follows. 

\begin{itemize}
    \item \textbf{TWAP (Time-Weighted Average Price) Order}:  
    A TWAP order aims to execute a large order evenly over a specified time period, minimizing market impact. The total quantity is divided into equal-sized smaller trades submitted at regular time intervals.

    \item \textbf{VWAP (Volume-Weighted Average Price) Order}:  
    A VWAP order seeks to match the market's trading volume profile. The execution quantity is proportional to the observed trading volume over time, allowing larger portions of the order to be executed when market activity is higher, thereby reducing slippage.

    \item \textbf{Iceberg Order}:  
    An iceberg order hides the true order size by only displaying a small portion of the total quantity on LOB. As the visible portion is filled, additional hidden shares are revealed, allowing traders to conceal their full trading intentions and limit market impacts.
\end{itemize}

We simulate this new FMS for a 2-hour trading period, setting the iceberg order quantity at 20,000 units with a visible portion of 100 units. The execution frequency for both TWAP and VWAP orders is set to once per minute. Agents are configured to execute trades in both the buy and sell directions, respectively. Additionally, the execution price is fixed at $\$$99.80.
From Table \ref{tab:advanced_orders}, it can be observed that our model demonstrates robustness across different advanced order types, highlighting its effectiveness in capturing the latent representation of limit order book series.

\begin{figure}[!t]
\centering
\includegraphics[width=3.5in,height=2.2in]{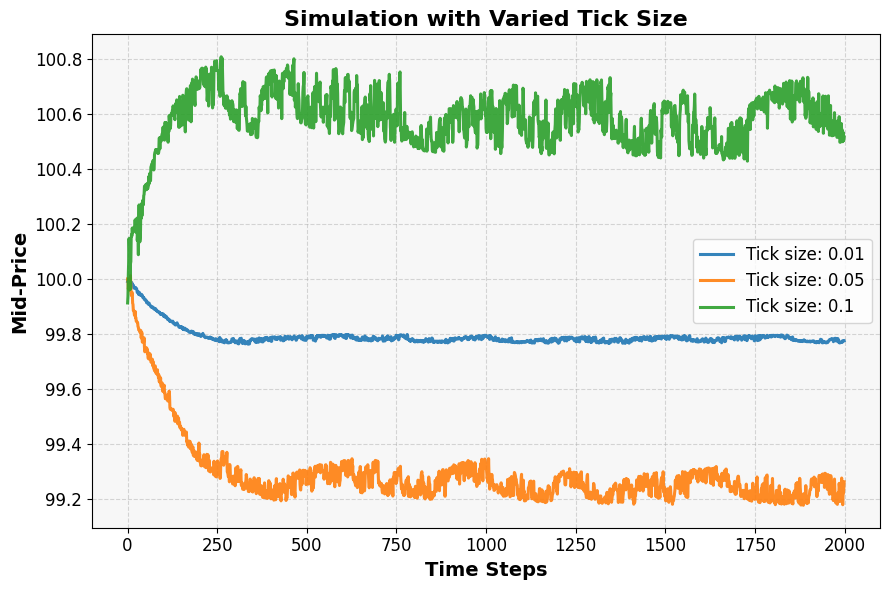}
\caption{{The mid-price curves with different tick sizes (\$).}}
\label{fig:Ticksize_window_100}
\end{figure}

Moreover, we evaluate the reconstruction erros of the compared networks with synthetic data under 3 tick sizes of $\$$0.01, $\$$0.05, and $\$$0.1, simulating both fine-grained and coarse-grained price discretization. We show that the varied tick sizes influence the mid-price trends significantly (Fig. \ref{fig:Ticksize_window_100}), while the reconstructed errors still verify the advantages of SimLOB on representation learning.
As shown in Table \ref{tab:ticksize_comparison_v2_transposed}, SimLOB-based variants consistently achieve the top-2 errors, indicating strong robustness to the price discretization. 



\begin{table*}[bht]
\centering
\caption{Reconstruction errors under different advanced order types.}
\label{tab:advanced_orders}
\begin{tabular}{c||c|c|c|c|c|c|c}
\hline\hline
\textbf{Order Type} & \textbf{SimLOB2} & \textbf{SimLOB8} & \textbf{CNN1} & \textbf{CNN2} & \textbf{Linear} & \textbf{DeepLOB} & \textbf{TransLOB} \\\hline\hline
\multirow{2}{*}{ICEBERG} 
& Buy \underline{0.0074} & \textbf{0.0068} & 0.0353 & 0.0145 & 0.0116 & 0.0341 & 0.0808 \\
& Sell \textbf{0.0152}    & \underline{0.0165} & 0.1375 & 0.0481 & 0.0435 & 0.1229 & 0.3081 \\\hline
\multirow{2}{*}{VWAP} 
& Buy \textbf{0.0156} & \underline{0.0170} & 0.1374 & 0.0486 & 0.0442 & 0.1230 & 0.3079 \\
& Sell \textbf{0.0158} & \underline{0.0172} & 0.1374 & 0.0488 & 0.0442 & 0.1230 & 0.3079 \\\hline
\multirow{2}{*}{TWAP} 
& Buy \textbf{0.0156} & \underline{0.0170} & 0.1372 & 0.0485 & 0.0443 & 0.1230 & 0.3080 \\
& Sell \textbf{0.0152} & \underline{0.0165} & 0.1374 & 0.0482 & 0.0437 & 0.1231 & 0.3081 \\\hline\hline
\end{tabular}
\end{table*}

\begin{table}[htbp]
\centering
\caption{Reconstruction Errors under Different Tick Sizes ($\$$).}
\label{tab:ticksize_comparison_v2_transposed}
\begin{tabular}{l||ccc}
\hline\hline
\textbf{Model} & \textbf{Tick = 0.01} & \textbf{Tick = 0.05} & \textbf{Tick = 0.10} \\
\hline\hline
SimLOB2   & \textbf{0.0036} & \underline{0.0035} & \textbf{0.0037} \\
SimLOB8   & \underline{0.0040} & \textbf{0.0031} & \textbf{0.0037} \\
MLP    & 0.0047          & 0.0047          & 0.0059 \\
CNN1      & 0.0134          & 0.0149          & 0.0116 \\
CNN2      & 0.0051          & 0.0063          & 0.0045 \\
TransLOB  & 0.0120          & 0.1226          & 0.0113 \\
\hline\hline
\end{tabular}
\end{table}

\subsection{Group 3 - Interpretability Analysis of SimLOB}

Interpretability is essential for machine learning models in finance. To evaluate the interpretability of SimLOB, we study the parameters and mechanisms of the FCN$_1$ and Transformer blocks.

The linear layer in FCN$_1$ has a weight matrix of size $40 \times 256$, mapping 40 original LOB features at each time step to 256 latent features. By averaging the weights across the first dimension (40 original features), we obtain a 256-dimensional vector representing the average importance of each latent feature. Averaging the weights across the second dimension (256 latent features) yields a 40-dimensional vector indicating the contribution of each original LOB feature. 

\begin{figure}[b]
\centering
\includegraphics[width=3.6in,height=3in]{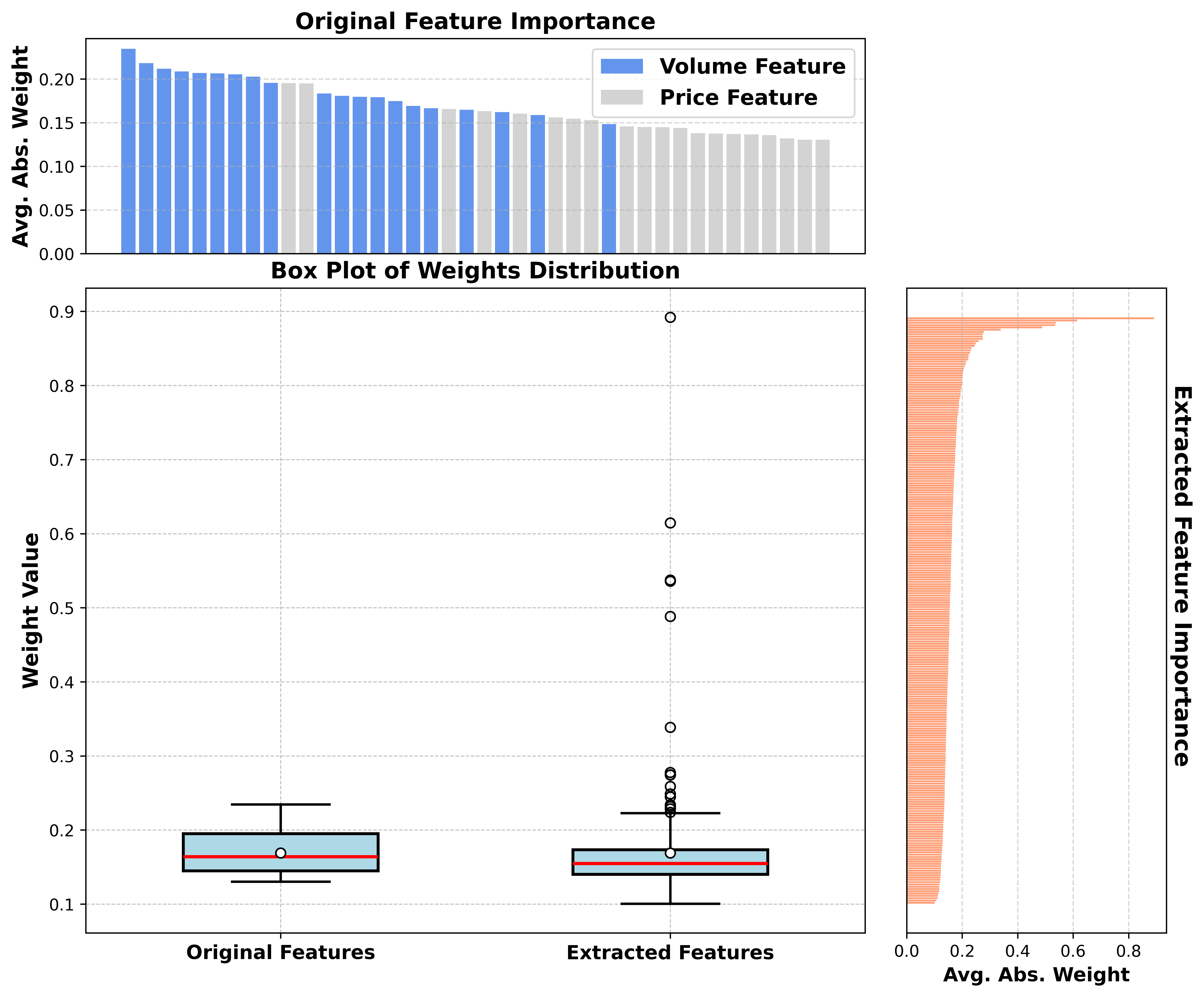}
\caption{{Visualization of feature importance and weight redistribution through the linear projection layer.}}
\label{fig:Visualization_weight_distribution}
\end{figure}

\begin{figure}[t]
\centering
\includegraphics[width=3.5in,height=4in]{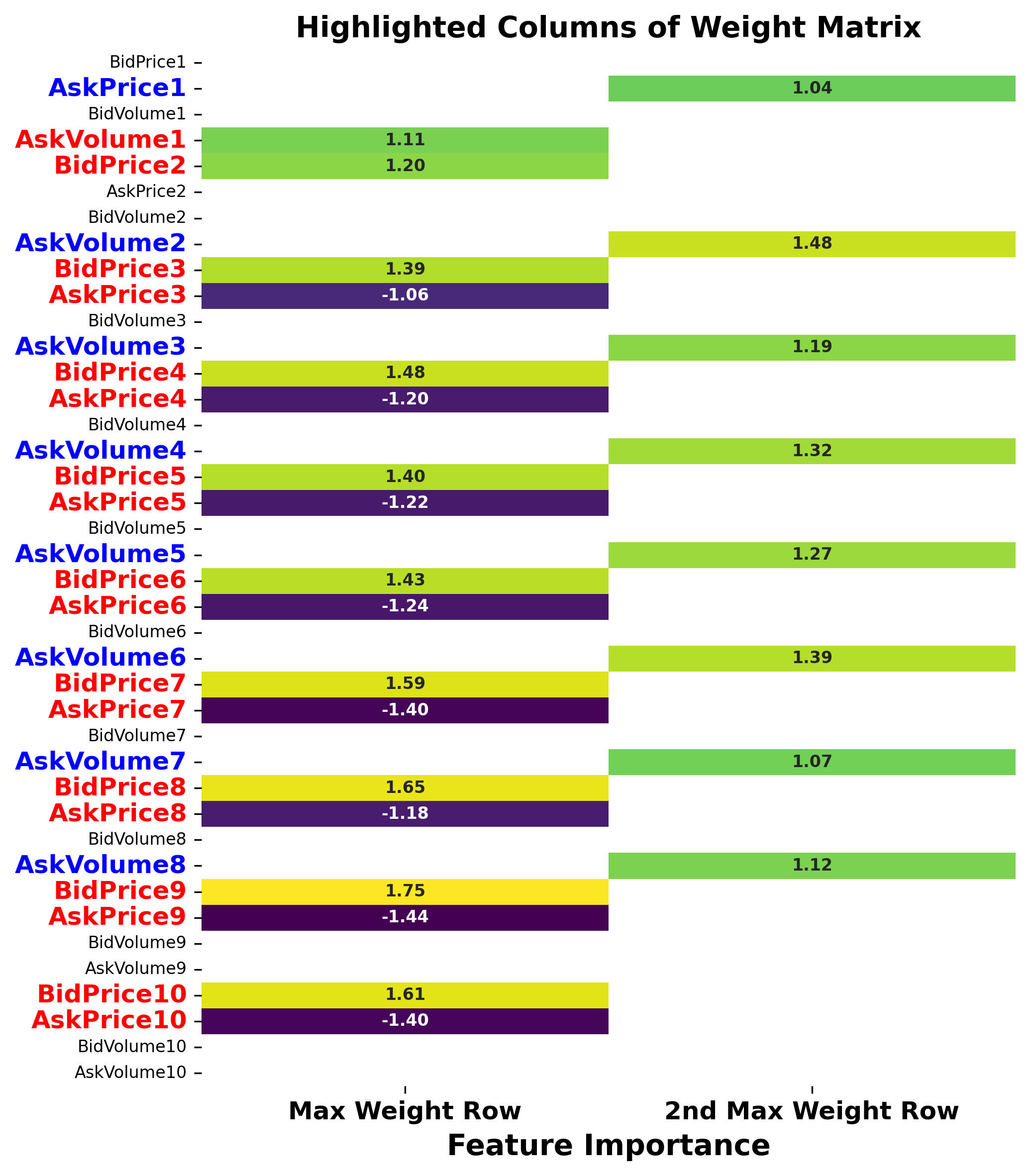}
\caption{{Visualization of the two most important latent features in the linear projection layer.}}
\label{fig:Visualization_2_latent}
\end{figure}

Fig. \ref{fig:Visualization_weight_distribution} visualizes the weight distribution in the linear projection layer. The central box plot displays the distribution of absolute weight values for both the original and extracted features. The top and right subplots present the average absolute weights for the original and extracted features, respectively. Volume and price features are distinguished using different colors. From the figure, we observe that the weights across the original 40 LOB features are relatively uniform, with a varying range of 0.10. After processing by FCN$_1$, the weights become more varied, with a range of 0.79, indicating a clear redistribution and differentiation in feature importance.

One of the key distinctions between LOB and mid-price is the inclusion of volume information. The Fig. \ref{fig:Visualization_weight_distribution} also shows that volume-related features rank relatively high in terms of the weights importance, with an average rank of 11.65 among all original LOB features. This suggests that the model effectively leverages volume signals, underscoring their critical contribution to prediction performance.

After processing by FCN$_1$, we identify the feature with the large absolute weights. After filtering out features with absolute weights less than 1, we surprisingly find that this dominant feature carries financial significance, i.e., it can be approximately expressed as $\sum_{i=3}^{10} BidPrice_i - 0.82 \times AskPrice_i$, reflecting an adjusted bid-ask spread.
Similarly, the second most important feature can be interpreted as an adjusted average ask volume and can be approximately expressed as $1.26 \times \sum_{i=2}^8 AskVolume_i$.

\begin{figure}[!t]
\centering
\includegraphics[width=3.5in,height=1.5in]{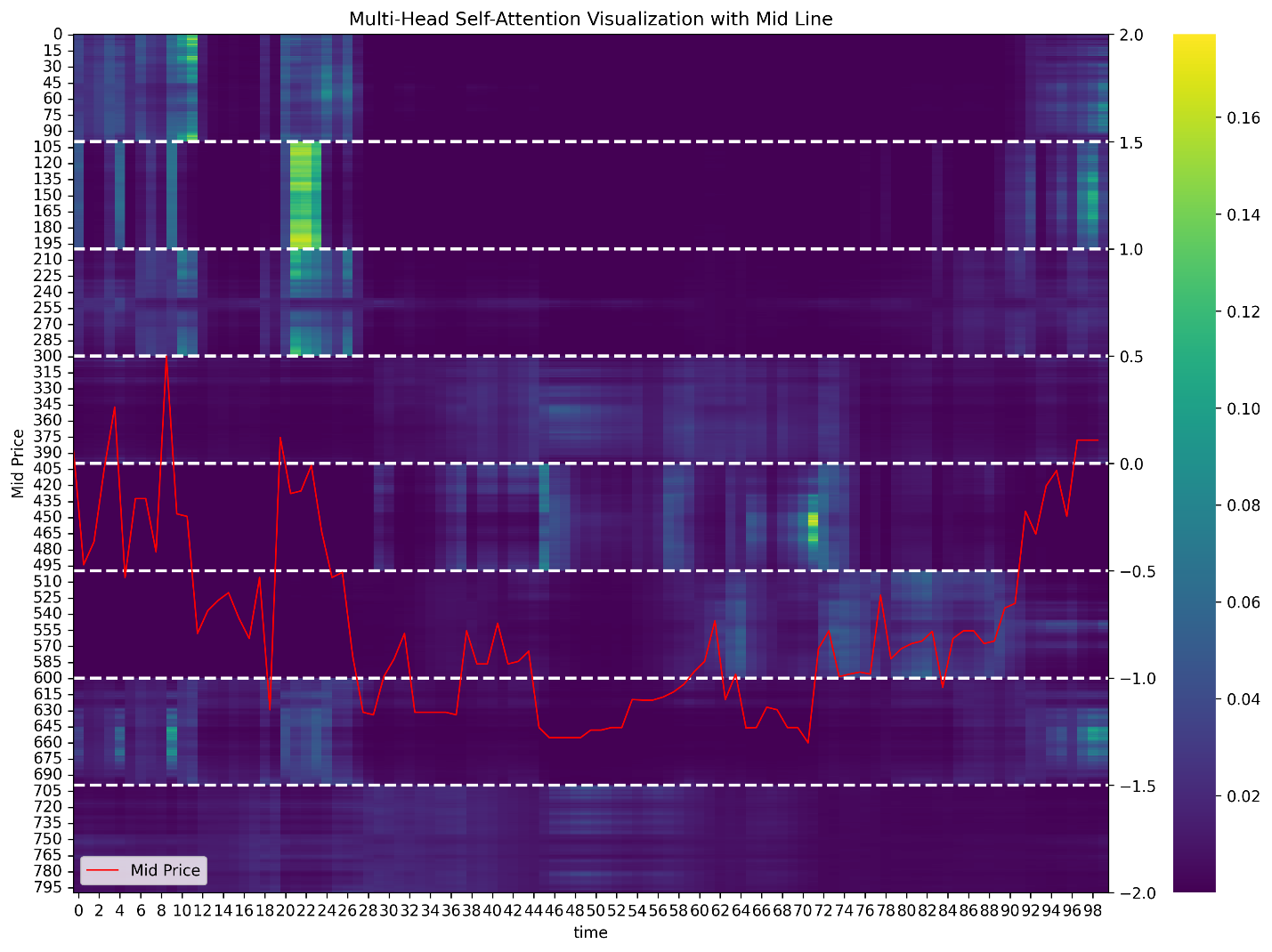}
\caption{{Multi-head self-attention visualization with mid price.}}
\label{fig:attention weight}
\end{figure}

To demonstrate the effectiveness of the Transformer blocks in SimLOB, we visualized the multi-head self-attention weights, as shown in the Fig. \ref{fig:attention weight}. This visualization directly addresses the concern regarding the role of attention in capturing the temporal patterns of LOB series. In this figure, the horizontal axis represents the timestamps of 100 consecutive LOB snapshots, illustrating the temporal progression of the data. For the vertical axis, we structured the vertical axis into eight blocks, each corresponding to one of the attention heads in our transformer model. Each block visualizes how a specific timestamp (query) attends to other timestamps (keys) based on the computed attention weights. We calculate the attention weight by computing the dot product between the query (Q) and key (K) vectors, scaling, and applying the softmax function as follows:
\begin{equation}
    \text{Attention}(Q, K, V) = \text{softmax}\left(\frac{QK^T}{\sqrt{d_k}}\right) V
\end{equation}

The color intensity directly indicates the weight magnitude, where lighter colors represent higher attention weights, indicating stronger relevance between timestamps, while darker colors represent weaker attention, indicating minimal interaction. On the right vertical axis, we plot the mid-price trend as a red line to illustrate the evolution of mid-price over the same period.
The visualization clearly shows that the attention weights are not uniformly distributed, exhibit structured patterns across time. We observe that some timestamps consistently attract higher attention weights, particularly around critical mid-price changes, like suddenly peaking or slowing. This concentration of attention indicates that the model effectively learns to focus on significant timestamps where mid-price fluctuations are most evident. There is a clear correlation between high attention weights and periods with significant mid-price variations. This alignment demonstrates that the model effectively identifies crucial historical contexts associated with sudden price shifts, reinforcing the utility of the attention mechanism in tracking LOB dynamics.

Besides, different attention heads capture different temporal patterns. Each attention head exhibits distinct focuses throughout the time series, indicating that different heads capture unique aspects of temporal dependencies. Some heads show attention at earlier time steps, indicating their ability to capture information in initial signals. Some heads distribute their attention evenly, indicating their ability to learn long-range dependencies and global attention. These diverse focuses across heads demonstrates the model’s ability to learn both local and global patterns, thereby enhancing its ability to understand complex temporal structures.

\begin{table}[b]
\centering
\caption{Ablation Studies of SimLOB on the Three Blocks}
\begin{tabular}{p{3cm}||p{3cm}}
\hline\hline
\textbf{Networks} & \textbf{Reconstructed Errors} \\
\hline\hline
SimLOB-v1 & $0.4581 \pm 4.9568$ \\
SimLOB-v2 & $0.0433 \pm 0.0164$ \\
SimLOB-v3 & $0.1446 \pm 0.0218$ \\
TransLOB & $0.4650 \pm 4.9561$ \\
SimLOB & \textbf{$0.0273 \pm 0.0089$} \\
\hline\hline
\end{tabular}
\label{table:Ablation Study and Comparison with TransLOB}
\end{table}

\subsection{Group 4 – Sensitivity and Ablation Analysis of SimLOB}
\label{sec:group4_sensitivity}
As both SimLOB and TransLOB utilize Transformer blocks, they are empirically compared in an ablation manner to demonstrate the effectiveness of the proposed three components of SimLOB: the fully connected feature extraction block, the transformer block stack, and the dimension reduction block. Comparatively, TransLOB also contains three components: a CNN-based feature extraction block, a transformer block, and a linear-layer output block. By replacing the individual blocks of SimLOB with their corresponding components from TransLOB, we created three variants of SimLOB. For clarity, SimLOB-v1 denotes the SimLOB variant whose first block is substituted by the first block of TransLOB. The same rule of creating SimLOB variants applies to SimLOB-v2 and SimLOB-v3.

As seen in Table \ref{table:Ablation Study and Comparison with TransLOB}, with CNN, both SimLOB-v1 and TransLOB exhibit similarly poor performance. The comparisons between SimLOB-v1 and SimLOB highlights that although CNN has been the mostly popular network in LOB literature, they are not well-suited for LOB representation learning. Furthermore, the results of SimLOB-v1 and SimLOB-v3 emphasize the importance of both the fully connected feature extraction block and the fully connected dimension reduction block for achieving optimal performance.

\begin{figure}[t]
\centering
\includegraphics[width=3.6in,height=1.6in]{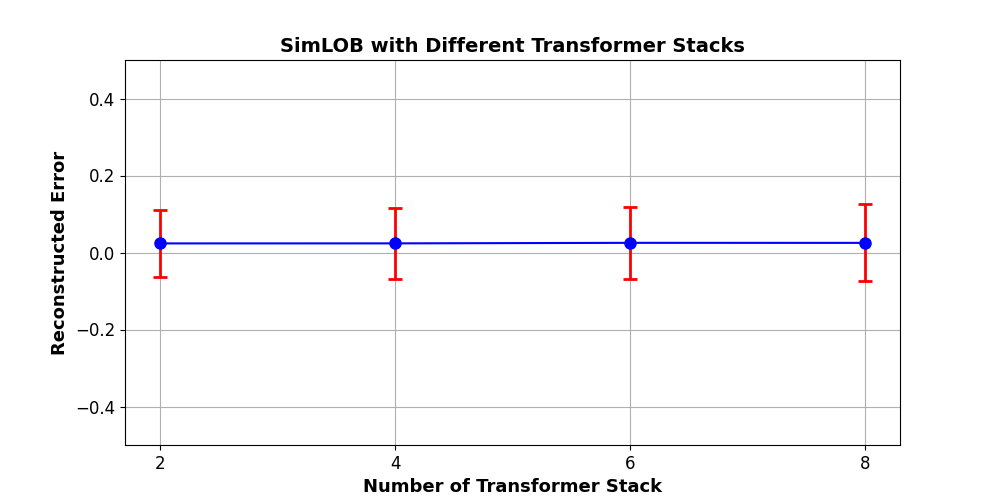}
\caption{Impacts of the number of Transformers stack on the reconstruction errors.}
\label{fig:number of Transformer Stack}
\end{figure}

\begin{table}[b]
\centering
\caption{Impacts of Different Representation Vector Lengths}
\begin{tabular}{p{3cm}||p{3cm}} 
\hline\hline
\textbf{Networks} & \textbf{Reconstruction Errors} \\
\hline\hline
$\mathrm{SimLOB}_{\tilde{\tau}=64}$ & $0.0498 \pm 0.1200$ \\
$\mathrm{SimLOB}_{\tilde{\tau}=128}$ & $0.0250 \pm 0.0997$ \\
$\mathrm{SimLOB}_{\tilde{\tau}=256}$ & $0.0182 \pm 0.1056$ \\
$\mathrm{SimLOB}_{\tilde{\tau}=512}$ & $0.0176 \pm 0.0844$ \\
\hline\hline
\end{tabular}
\label{table:On the Length of Representation Vector}
\end{table}

\begin{table}[t]
\setlength{\tabcolsep}{2.5pt} 
\centering
\caption{Calibration Errors for Different Representation Vector Lengths}
\begin{tabular}{l||cccc}\hline\hline
&$\mathrm{SimLOB}_{\tilde{\tau}=64}$   & $\mathrm{SimLOB}_{\tilde{\tau}=128}$  & $\mathrm{SimLOB}_{\tilde{\tau}=256}$  & $\mathrm{SimLOB}_{\tilde{\tau}=512}$  \\\hline\hline
data1 &3.53E03          & 1.27E03          & \textbf{8.68E02} & 1.03E03          \\
data2 &7.71E03          & 1.30E03          & 6.88E03          & \textbf{1.25E03} \\
data3 &5.36E02          & \textbf{4.35E02}          & 4.75E02 & 4.79E02          \\
data4 &1.86E03          & 5.51E02          & 5.26E02          & \textbf{4.10E02} \\
data5 &1.62E03          & \textbf{1.20E03} & 2.49E03          & 1.38E03          \\
data6 &\textbf{5.90E03} & 8.58E03          & 7.23E04          & 8.39E03          \\
data7 &2.08E05          & 1.49E06          & 1.65E05          & \textbf{1.60E05} \\
data8 &1.72E03          & \textbf{1.07E03} & 1.23E03          & 1.11E03          \\
data9 &1.51E04          & \textbf{1.13E04} & 2.64E04          & 4.01E04          \\
data10 &5.82E03          & \textbf{9.83E02} & 1.35E04          & 7.39E03         \\\hline\hline
\end{tabular}
\label{table:The Calibration errors of different representation vectors}
\end{table}

To analyze the architecture settings in SimLOB, we mainly focus on the number of Transformer Stack and the length of the representation vector. For each parameter, multiple SimLOB variants are tested with different configurations.




\textbf{The number of Transformer Stack.} 
Fig. \ref{fig:number of Transformer Stack} shows the reconstruction errors of SimLOB with $L=2,4,6,8$ Transformer stacks. As the result shows, increasing the number of transformer stacks has minimal impact on the performance. The reconstruction errors across 0.2 million testing samples for all four variants remained consistently around 0.025, with a standard deviation of approximately 0.095. Given that adding an additional Transformer stack introduces roughly 1.5 million more parameters, $L=2$ is selected as the default setting for SimLOB to balance performance and efficiency. This result indicates that while the Transformer is useful, it is not the most critical component.

\textbf{The Length of Representation Vector.} 
We conduct the sensitive analysis on the length of representation vector $\tilde{\tau}$ with different choices like 64, 128, 256, and 512, denoted as $\mathrm{SimLOB}_{\tilde{\tau}=64}$, $\mathrm{SimLOB}_{\tilde{\tau}=128}$, $\mathrm{SimLOB}_{\tilde{\tau}=256}$ and $\mathrm{SimLOB}_{\tilde{\tau}=512}$. SimLOB is trained with those settings separately and tested on the 0.2 million testing samples.

The reconstruction errors, listed in Table \ref{table:On the Length of Representation Vector}, are shown that longer representation vectors leads to better reconstruction errors, suggesting larger $\tilde{\tau}$ for representation learning.  The intuition is that the longer the latent vector is, the more detailed information of the time series LOB can be embedded. However, as Table \ref{table:On the Length of Representation Vector} shows, this advantage decreases quickly beyond $\tilde{\tau}=128$.
At the same time, larger lengths increases complexity of downstream tasks without necessarily improving calibration performance. This can be seen in Table \ref{table:The Calibration errors of different representation vectors}, where $\tilde{\tau}=128$ achieves the best results in 5 out of 10 instances. 
Thus, by balancing the reconstruction errors and the complexity, $\tilde{\tau}=128$ is selected as the default setting for SimLOB, which has already outperformed the SOTA.

\section{Conclusions and Future Works}
This paper proposes to learn the vectorized representations of the LOB data, the fundamental dataset in the financial market, using an autoencoder framework. The latent vector is taken as the learned representation.
Although increasing research works focused on analyzing the LOB with neural networks, they did not use the autoencoder framework or  explicitly learned the vectorized representations.
This work studies 8 SOTA neural architectures for LOB, and discusses that the commonly used convolution layers are not effective for preserving the specific properties of LOB.
Based on that, a novel neural architecture is proposed with three components: a fully connected network for extracting the features of price precedence from LOB, a stacked Transformers for capturing the auto-correlation, and another fully connected network for reducing the dimensionality of the latent vector.
Empirical studies verify the advantages of the proposed network in terms of the reconstruction errors against existing neural networks for LOB.
Moreover, the representation learning shows a positive correlation with the downstream FMS calibration performance.
Empirical results support that FMS models should be calibrated using compact LOB representations rather than relying solely on mid-prices or raw LOB data. 
The proposed SimLOB is also verified to be able to generalize to real data from different markets as well as with advanced market settings, like order types and tick sizes.
Furthermore, it is visualized that SimLOB is interpretable for its representation ability in both cross-features and temporal correlations.
At last, a surprising finding is that the mostly adopted CNN architecture is actually not suitable for LOB.

In our current setup, the model demonstrates fast inference speed—reconstructing a duration of 5 minutes LOB data (100 time steps) takes approximately 0.18 seconds. In this case, we are able to retrain on a full trading day’s data (around 4800 time steps a stock and 5000 stocks in China market for example) within 5 minutes, making it feasible to update the model very quickly.
This level of efficiency enables near-real-time deployment, provided adequate computational resources. Furthermore, the retraining pipeline allows for frequent adaptation to evolving market conditions, enhancing the system's resilience and practical utility for automated calibration in live trading environments.

In the future, we plan to develop a penalty-based loss function to guide the model in learning more accurate LOB patterns. Additionally, we may consider incorporating penalties based on discrepancies from stylized facts, which would further improve the model's adherence to LOB structure.

\bibliographystyle{IEEEtran}
\bibliography{manuscript-simlob}

\end{document}